\documentstyle[pre,aps,amsfonts,amssymb,epsf]{revtex}

\newcommand{\be}{\begin{equation}}
\newcommand{\ee}{\end{equation}}
\newcommand{\bea}{\begin{eqnarray}}
\newcommand{\eea}{\end{eqnarray}}
\newcommand{\bfig}{\begin{figure}}
\newcommand{\efig}{\end{figure}}
\newcommand{\bl}{\begin{LARGE}}
\newcommand{\el}{\end{LARGE}}

\newcommand{\rb}{\mbox{${\bf r}$}}
\newcommand{\rlb}{\mbox{${\bf R}$}}
\newcommand{\fb}{\mbox{${\bf F}$}}
\newcommand{\fc}{\mbox{$F_{\rm c}$}}
\newcommand{\kt}{\mbox{$k_{\rm B} T$}}
\newcommand{\ktm}{k_{\rm B} T}
\newcommand{\ett}{\mbox{$\tilde{\eta}$}}
\newcommand{\uhat}{\mbox{${\bf u}$}}
\newcommand{\cale}{{\cal E}}
\newcommand{\hc}{\mbox{${\cal H}_{\rm C}$}}
\newcommand{\hcp}{\mbox{${\cal H}_{\rm C,pull}$}}
\newcommand{\call}{{\cal L}}
\newcommand{\mavg}{\mbox{$\overline{\langle m \rangle}$}}
\newcommand{\zt}{\mbox{$\tilde{Z}$}}
\newcommand{\ztm}{\tilde{Z}}
\newcommand{\phat}{\hat{P}}
\newcommand{\mtherm}{\mbox{$\langle m \rangle$}}
\newcommand{\mmin}{\mbox{$m_{\text{min}}$}}
\newcommand{\pineq}{\mbox{$P_{\text{ineq.}}$}}
\newcommand{\emin}{\mbox{$\cale_{\text{min}}$}}
\newcommand{\dm}{\mbox{$m_{\text{jump}}$}}
\newcommand{\dmm}{m_{\text{jump}}}
\newcommand{\pr}{\text{Prob}}
\newcommand{\de}{\mbox{$\cale_{\text{jump}}$}}
\newcommand{\pj}{\mbox{$P_{\text{jump}}$}}
\newcommand{\epat}{\mbox{$\varepsilon_{\text{AT}}$}}
\newcommand{\epgc}{\mbox{$\varepsilon_{\text{GC}}$}}
\newcommand{\fss}{\mbox{$F_{\text{ss}}$}}
\newcommand{\zext}{\mbox{${\cal Z}$}}
\newcommand{\ravg}{\mbox{$\overline{\langle r \rangle}$}}
\newcommand{\rtherm}{\mbox{$\langle r \rangle$}}
\newcommand{\favg}{\mbox{$\overline{\langle F \rangle}$}}
\newcommand{\df}{\mbox{$\delta \! {\cal F}$}}
\newcommand{\delforce}{\mbox{$\delta \! F$}}
\newcommand{\tss}{\mbox{$\tau_{\text{ss}}$}}
\newcommand{\tend}{\mbox{$\tau_{\text{end}}$}}
\newcommand{\tbulk}{\mbox{$\tau_{\text{bulk}}$}}
\newcommand{\trot}{\mbox{$\tau_{\text{rot}}$}}
\newcommand{\rotmob}{\mbox{$\mu_{\text{rot}}$}}

\newcommand{\parenref}[1]{(\ref{#1})}

\begin{document}

\title{Single Molecule Statistics and the Polynucleotide Unzipping Transition}
\author{David K. Lubensky~\cite{my-email} and David R. Nelson~\cite{nelson-email}}
\address{Department of Physics, Harvard University \\ Cambridge MA  02138}

\maketitle

\begin{abstract}
We present an extensive theoretical investigation of the mechanical
unzipping of double-stranded DNA under the influence of an applied
force.  In the limit of long polymers, there is a thermodynamic {\em
unzipping} transition at a critical force value of order 10
$\text{pN}$, with different critical behavior for homopolymers and for
random heteropolymers. We extend results on the disorder-averaged
behavior of DNA's with random sequences~\protect\cite{me-prl} to the
more experimentally accessible problem of unzipping a single DNA
molecule. As the applied force approaches the critical value, the
double-stranded DNA unravels in a series of discrete,
sequence-dependent steps that allow it to reach successively deeper
energy minima.  Plots of extension versus force thus take the striking
form of a series of plateaus separated by sharp jumps.  Similar
qualitative features should reappear in micromanipulation experiments
on proteins and on folded RNA molecules.  Despite their unusual form,
the extension versus force curves for single molecules still reveal
remnants of the disorder-averaged critical behavior.  Above the
transition, the dynamics of the unzipping fork is related to that of a
particle diffusing in a random force field; anomalous,
disorder-dominated behavior is expected until the applied force
exceeds the critical value for unzipping by roughly 5 pN.
\end{abstract}

\section{Introduction}
Over the past decade, the experimental repertoire of biophysicists and
structural biologists has expanded to include some remarkable
micromanipulation techniques.  These single molecule methods are
a natural complement to more traditional scattering and spectroscopic
measurements: Although they cannot ascertain structures at atomic
resolution, they do give important information about the organization of disordered or strongly-fluctuating systems, and they yield
valuable estimates of the forces and energies that
stabilize a given structure.  Moreover, micromanipulation experiments
on single molecules
open a window into a rich and largely unexplored set of physical
phenomena.  One can now measure entire distributions of molecular
properties, without the requirement for averaging over a macroscopic
sample.  Not only does the wealth of resulting data allow more
stringent tests of ideas originally developed for macroscopic systems,
it also has the potential to reveal entirely new behavior that was not
discernible in aggregate results on heterogeneous populations of
molecules~\cite{cluzel,chu,liphardt}.  In this paper, we study an
example of a system---the {\em unzipping} of double-stranded DNA
(dsDNA)---that shows exactly such novel response on the single
molecule level.  Our results are also directly applicable to the
unzipping of a single RNA hairpin, and similar ideas can be applied to
the force-induced denaturation of RNA's with more complicated
secondary structures~\cite{liphardt} and even to the stretching of folded proteins~\cite{protein-pull}.


In the DNA unzipping problem, the two single strands of
a double-stranded DNA molecule with a randomly chosen base sequence are pulled
apart under the influence of a constant force (Fig.~\ref{fig1}).  In
addition to providing a surprisingly good description of protein-coding
DNA~\cite{dna-stat}, the assumption of a random sequence gives us an
analytically tractable model; its solution then allows
us to gain insight into a much broader class of systems.  DNA
unzipping thus serves as a model problem to illuminate the effect of
sequence variation on a micromechanical experiment.

In a previous brief communication~\cite{me-prl},
we showed that the {\em average} extension versus force curve of an
ensemble of random heteropolymers is markedly different from the
corresponding curve for a homopolymer.  Here, we move beyond averages
over many different random sequences to examine the unzipping of a
{\em single} dsDNA molecule.  Interesting qualitative lessons emerge.
Whereas a homopolymer gains considerable entropy by opening in
response to a constant force, a heteropolymer unzips primarily for
energetic reasons.  In fact, the unzipping process is dominated by the
presence of deep energy minima and is only mildly perturbed by thermal
fluctuations.  At any given applied force, the system will sit in the
deepest available minimum; because the location of the minimum varies
discontinuously with the applied force, the number of bases opened
will show sharp jumps at certain force values.  Moreover, the energy
landscape is determined by the polymer's sequence, so the
force-extension curve will be strongly sequence-dependent.

A number of theorists have recently addressed aspects of dsDNA
unzipping~\cite{siggia,peyrard,bhatta,sebastian,maritan1,maritan2,maren-bhatta-seq,cocco-marko};
the mechanical properties of a single-stranded polynucleotide that can
pair with itself have also received considerable
attention~\cite{mezard,zhou1,zhou2,terry-uli}.  With a few
exceptions~\cite{maren-bhatta-seq,terry-uli}, however, this work has been restricted to
the study of homopolymers, and thus does not overlap directly with the
results presented here.  

Although our model is chosen more for its
simplicity than for a clear correspondence to a particular experiment
in the literature, several related experiments have nonetheless been
performed.  Early studies by Lee and coworkers~\cite{lee} were
followed by the ground-breaking work of Essevaz-Roulet, Bockelmann,
and Heslot~\cite{heslot}, who demonstrated the feasibility of
mechanically denaturing single dsDNA molecules, and showed that many features of their
results could be understood using equilibrium statistical mechanics.
Subsequently, similar experiments have been performed using an atomic
force microscope~\cite{gaub,strunz}.  In contrast to our calculations,
this work was done in an ensemble in which the positions of the two
single-stranded ends are held fixed while an average force is
measured.  Because of subtleties associated with the statistical
mechanics of single molecule systems, this {\em constant
extension} ensemble is not equivalent in the usual sense to our {\em
constant force} ensemble; the connection between the two will be
discussed in more detail in section~\ref{const-ext-sect}.  More
recently, Liphardt and coworkers have mechanically unfolded several
different short RNA molecules related to a domain of the {\em
Tetrahymena thermophila} ribozyme~\cite{liphardt}.  Here, a bead
tethered to a force-measuring optical trap was used both to impose an
extension and, with feedback, to
monitor extension at fixed force---precisely the situation of interest
in this paper.  Alternatively, a constant force could be applied
directly using a magnetic bead in a constant magnetic field
gradient~\cite{wirtz-amblard}.

In the remainder of this paper, we first, in Section~\ref{model-sect}, describe in more detail the
phase diagram of polynucleotide duplexes and show how a coarse-grained
model of the unzipping transition can be derived from more microscopic
descriptions of dsDNA.  This model, which will form the basis of all
subsequent calculations, is summarized in Eqs.~\parenref{cale-defn}
through~\parenref{Z-defn}.  For the purposes of comparison, we
derive in Section~\ref{non-rand-sect} some results on the unzipping of homopolymeric dsDNA.
Section~\ref{disorder-average} revisits in more detail the
disorder-averaged force-extension curve examined in~\cite{me-prl}.
The bulk of our new results on single-molecule unzipping appear in
section~\ref{single-mol}.  We show that the equilibrium extension versus force curve
of a single dsDNA molecule consists of a series of long plateaus
followed by large jumps, and we derive a statistical description of
this striking behavior.  We also demonstrate that, despite its choppy
appearance, such a curve contains hidden signatures of the smooth
disorder-averaged behavior.  Subsequent sections consider the
relationship between the conjugate constant force and constant
extension ensembles (Sec.~\ref{const-ext-sect}) and give a brief overview of the dynamics of
unzipping (Sec.~\ref{dynamics}).   We point out that polynucleotide
unzipping provides an experimental realization of the famous Sinai
problem of thermally activated diffusion in a quenched random force
field~\cite{sinai,bouch-geo}.  Anomalous, quasi-localized dynamics
persist up to roughly $5 \text{pN}$ above the unzipping transition. 
 Finally,
in Section~\ref{conclusions}, we discuss the implications of DNA
unzipping for
micromanipulation experiments on more complicated systems.  The
appendix gives a brief description of the numerical methods used to
generate results discussed in the body of the paper.


\section{The Model}
\label{model-sect}

Figure~\ref{fig1} depicts the situation studied in this paper: One of the
single strands from a double-stranded DNA
molecule is attached to a glass slide, and the other to a bead on
which a constant force \fb\ is exerted.  \fb\ could be created, for
example, with magnetic tweezers, which have been used to exert
constant piconewton-scale forces over hundreds of
microns~\cite{wirtz-amblard}.  Optical tweezers or atomic force
microscopes (AFM) with appropriate feedback can create a similar
effect~\cite{liphardt,block-review}.  As a result of the applied force, the DNA
partially ``unzips'', breaking $m$ bonds.  As long as the
force-elongation curve of the liberated single-stranded DNA is known,
$m$ can be related to the distance $r$
between the ends of the two single strands, which is easily
measured.  Our main goal
is to understand how the {\em equilibrium} ensemble average $\langle m
\rangle$ (where the angle brackets indicate an average over thermal
noise) depends on $F$ and on the base sequence of the DNA strand.

In certain limiting cases, the
dependence of $m$ on $F$ is easy to understand.  One might expect
that at large enough forces the dsDNA will unzip completely, whereas
for very small forces at most a few bases will open.  We show
below that these two regimes are separated by a sharp first order
phase transition.  Below the critical force \fc, only a finite number
of bases at the end of the double strand are pulled open; in the
thermodynamic limit of an infinitely long DNA molecule, the pulling
force thus has no effect on the {\em fraction} of open bases, which
remains very small in physiological conditions.  Above \fc, the entire
molecule unzips, and the fraction of open bases jumps discontinuously
to one.  This phase diagram is sketched in the inset to
Figure~\ref{fig1}.  As $F$ approaches \fc\ from below, the number $m$
of unzipped bases at the end of the molecule diverges.  Because this
divergence is entirely a surface phenomenon, the unzipping transition
can be thought of as the one-dimensional analog of a continuous
wetting transition~\cite{wetting-reviews}.

The effect of base sequence on the force-elongation curve is less
straightforward.  We can gain some insight into the role
of a variable sequence by considering the problem of unzipping a DNA
molecule where each successive base is chosen at random, with at most
short-ranged correlations between bases.
Although the sequence of protein-coding DNA is certainly not
random in any strict sense, it nonetheless appears to many statistical
criteria to fit this description (up to
a length scale set by the sequence's mosaic
structure)~\cite{dna-stat}.  Deviations from randomness that
escape these tests presumably involve fairly subtle multi-point
correlations.  Although the structure of the protein for which the DNA
codes is likely to depend on such correlations,
the mechanical denaturation
of the DNA itself, which depends only on the cumulative energy cost of
opening $m$ bases, should be relatively insensitive to them.
Simulations of the more complicated problem of pulling on folded RNA's
have shown good agreement with the predictions of a random model~\cite{terry-uli}.
It is thus reasonable, at least as a first approximation, to take the
DNA sequence being unzipped to be random and uncorrelated.
In the remainder of this section, we develop a mathematical
description of the unzipping of such a DNA sequence by a constant force.

\subsection{Semi-Microscopic Models}
\label{semi-micro}

The bulk thermally-driven {\em melting} transition of dsDNA (see
Fig.~\ref{fig1}) can be described at varying levels of detail by a
number of models, all of which are expected to give the same universal
behavior on long enough length scales.  One popular choice is an
Ising-like description, in which a base pair is taken to be in one of
two discrete states---open or closed.  By convention, the free energy
of an unconstrained base pair in the open state is set to zero.  A
melted stretch of single-stranded DNA flanked by two unmelted regions
must form a closed loop, and a loop factor accounts for the loss of
entropy caused by this constraint~\cite{ising-melt}.  The Hamiltonian of a
semi-infinite strand can be written as a sum of (free) energies associated
with successive paired and unpaired regions:
\be
 {\cal H}_{\rm I} =  \sum_{i} \left\{ \left[
\sum_{n=c_i}^{o_i-1} \varepsilon_n \right] + 2J + f(c_{i+1} - o_i) \right\}
\label{hi}
\ee
Here base positions are indexed by $n \in \{0,1,2,\ldots\}$, and the
$i^{\rm th}$ closed and open sections start at base numbers $c_i$ and
$o_i$ respectively (see Fig.~\ref{ising-fig}).  Each base pair gains an energy $\varepsilon_i$ from
being closed; sequence dependent stacking interactions can be included
by adding an additional energy $\varepsilon_{n,n+1}$~\cite{pol-scher}.  For the case of a
random DNA sequence, the $\varepsilon_i$ are independent random
variables.  The energy $2 J$ per open section gives the energetic cost
of initiating a melted region, and
$f(c_{i+1} - o_i) \propto \ln 2 (c_{i+1} - o_i)$ is the entropic penalty associated with forming a
closed loop of length $2(c_{i+1} - o_i)$.  If there are open bases at the
end of the molecule, before the first closed section, they are counted
as the zeroth open section and do not incur any loop penalty.  The
model's partition function is a sum over all possible opening and
closing points, $Z_{\rm I} =
\sum_{0 \leq c_1<o_1<\ldots<c_n<o_n\ldots} \exp(-{\cal H}_{\rm I}/\ktm)$.

Alternatively, some models of the melting transition are written in
terms of the position of each base in three-dimensional space~\cite{cont-melt}.  In the continuum
limit, the simplest such description of a dsDNA of finite length $N$ has the Hamiltonian
\be
{\cal H}_{\rm C} =  \int_0^{N} dn \left\{ \, \frac{\ktm d}{4 a b}
\left(\frac{d \rlb}{dn}\right)^2 + V_n[\rlb(n)] \, \right\} 
\label{hc}
\ee 
where $\rlb(n)$ is the relative displacement of the two single strands
at base pair $n$, $d$ is the spatial
dimension, $a$ is the backbone length of a chemical monomer along
a single strand, and $b$ is the Kuhn length of single-stranded DNA (see
Fig.~\ref{var-defn}); the factor of $1/a b$
appears instead of the more usual $1/b^2$~\cite{doi-ed} because $n$ indexes base pairs
rather than Kuhn segments.  We will usually be interested in the limit $N
\rightarrow \infty$ of a semi-infinite polymer, just as for the
Ising-like model.  By convention, $\rlb(n) = \bbox{0}$ when the
$n^{\text{th}}$ set of bases are paired.  Because we will be
especially interested in the distance between the ends of the two single
strands, it is useful to define the {\em extension}
\be
\rb \equiv \rlb(0) \; .
\ee
The first term in Eq.~\parenref{hc} describes the entropic elasticity
of the single strands~\cite{doi-ed} and thus has the same effect as
the loop factors in the Ising-like model.  The second term accounts
for the attractive interactions between the two single strands.
Coarse-grained over a number of bases, they are described by a
phenomenological potential energy term 
\be
V_n[\rlb(n)] = [1 + \ett(n)]
h[\rlb(n)] \; . \label{v_of_n}
\ee
  Here $h$ is a short-ranged attractive potential, and the variation with base sequence of the strength of the
attraction between strands is described by $\ett(n)$.  Standard
methods show that the continuum partition function $Z_{\rm C}(\rlb,N) \equiv \int {\cal
D}[\rlb'(n)] \exp(-{\cal H}_{\rm C}[\rlb']/\ktm)$ obeys an imaginary time
Schr\"{o}dinger equation~\cite{doi-ed}.

Either model can readily be extended to include a force pulling apart the
double-stranded molecule.  We first show explicitly how this can
be done neglecting long-ranged interactions (e.g. excluded
volume or base-pairing interactions) within the liberated single
strands.  Subsequently, we will argue that
including such effects will lead to only minor changes in
our
results near enough to the transition.  A constant force acting at the end of the DNA ($n=0$) to
separate the two single strands contributes an energy that is linear
in their separation.  In the case of the continuum model~\parenref{hc}, one
must thus add a term to the Hamiltonian of the form
\be
{\cal H}_{\rm C,pull}(\fb) = - \fb \cdot \rb = \int_0^N dn \, \fb
\cdot d \rlb/dn \; .
\label{hcpull}
\ee
In writing the second equality, we have neglected the effect of the other
end of the dsDNA at $N = \infty$ ; with a physical polymer of finite
length $N$, this approximation should be valid as long as the number of open bases
$m \lesssim N$, so that $\rlb(N) \approx 0$. 

Unlike the continuum model, the Ising-like model does not keep track
of the positions of the open bases.  We must thus take an alternative
view of the effect of an unzipping force.  The last equality of
equation~\parenref{hcpull} gives a hint of how to do this.  Suppose
that, as in Fig.~\ref{ising-fig}, the first
closed section of dsDNA starts at base $c_1$, so that $m = c_1$
bases are unzipped by the force.  In the discrete Ising-like model,
each liberated single strand can be described as a string of $m$ individual
monomers.  The $n^{\text{th}}$ such monomer contributes a displacement
$\uhat_n^{\text{a}}$ or $\uhat_n^{\text{b}}$ to the total end to end
distance of the single strand, where the
superscripts distinguish the two strands.  The energy of
unzipping is thus  $- \fb \cdot \rb =
- \sum_{n=0}^{m} \fb \cdot \uhat_n^{\text{a}} + \sum_{n=0}^{m}
\fb \cdot \uhat_n^{\text{b}}$.
Note that there is no reason to
extend the sums over $n$ to infinity; the positions of base pairs beyond the
first closed pair have no effect on the end to end distance $\rb
\equiv \rlb(0)$.
We would now like to trace over the $\uhat$'s to obtain a contribution
to the Hamiltonian that depends only on the number of open monomers $m = c_1$.  The precise result
will depend on the model used to describe the elastic properties of a
single-stranded monomer.  For any reasonable choice, however, the
traces over the different \uhat's must decouple, leading to a free
energy of the form $2 m g(F)$.  Here $g(F)$ is the change in free energy of a
single-stranded monomer caused by applying a tension $F$;
 by
definition, $g(0) = 0$.  Because the
monomers gain energy by aligning with the pulling
force, $g(F)$ decreases with increasing $F$.  For example,
the continuum model Hamiltonian~\parenref{hc} is quadratic in $d\rlb/dn$ and thus describes
a polymer that responds linearly to an arbitrarily large force.  Such
a Gaussian model results in a $g(F)$ that is quadratic in $F$:
\be
g(F) = -\frac{a}{b} \frac{F^2 b^2}{2 d \kt} \; . \; \; \; \; \;
\text{(Gaussian)} \label{g-G}
\ee
Similarly, for an inextensible freely jointed chain, one finds~\cite{busta}
\be
g(F) = - \frac{a}{b}\kt \ln \left[ \frac{\kt \sinh(F b/\kt)}{F b} \right] \; . \; \; \; \; \;
\text{(Freely Jointed)} \label{fjc}
\ee
In these equations, $a$ is again the backbone distance between bases; the
factor of $a/b$ is necessary because $g(F)$ is defined as the free
energy per chemical monomer, not per Kuhn length.  More generally, if
the force $\fss(x)$ exerted by the single-stranded polymer as a function
of the extension $x$ per base can be measured, then
\be
g(F) = \int_0^{x(F)} \fss(x') d\!x' - F x(F) = -\int_0^F x(F') d\!F'
\;, \label{g(F)-gen}
\ee
where $x(F)$ is the inverse function of $\fss(x)$.

Regardless of the exact form of $g(F)$, the effect of an unzipping
force can be included in the Ising-like model by adding a term to the
Hamiltonian~\parenref{hi} that gives the free energy of the unzipped
monomers under tension.  Because $m = c_1$, we have ${\cal H} = {\cal
H}_{\text{I}} + {\cal H}_{\text{I,pull}}$ with
\be
{\cal H}_{\rm I,pull}(F) = 2 c_1 g(F) \; .  
\ee 
Since $g(F) <
0$, this term favors increasing $c_1$, and thus unzipping
the dsDNA.

\subsection{Reduction to One Degree of Freedom}

Semi-microscopic models like those just discussed contain far more
detail than is necessary to describe the unzipping transition.
Our calculations would simplify if we could integrate out nonessential
degrees of freedom
to obtain a description that focuses on the number of unzipped bases
$m$.  The full partition function of the Ising-like model is a sum over all of
the closing and opening points $c_1,c_2,c_3,\ldots$ and
$o_1,o_2,o_3,\ldots$.  Among these parameters, the
only one that determines the number of bases that have been unzipped is
$c_1$.  Hence we focus on a constrained partition function with $c_1 = m
$ fixed
\be
Z_{\rm I}^0 \exp\left[-\frac{ \cale(m) }{\ktm} \right] \equiv
\sum_{m=c_1<o_1<c_2<o_2<\ldots<c_n<o_n<\ldots} \exp\left[ -\frac{{\cal
H}_{\rm I} + {\cal H}_{\rm I,pull}(F) }{\ktm} \right] \; ,
\ee
where the partition function $Z_{\rm I}^0 =
\sum_{0 = c_1<o_1<\ldots<c_n<o_n\ldots} \exp[-{\cal H}_{\rm I}/\ktm -
{\cal H}_{\rm I,pull}(F)/\kt]$ with $c_1$ constrained to be zero is included so that $\cale(0) = 0$.
This expression defines the function $\cale(m)$; it can be introduced in a similar manner in the continuum model, by
adding the constraint $\rlb(m) = \bbox{0}$ to the partition function  $Z_{\rm C}(\rlb,N) \equiv \int {\cal
D}[\rlb'(n)] \exp(-{\cal H}_{\rm C}/\ktm) - {\cal H}_{\text{C,pull}}/\ktm)$ and replacing the attractive
potential $V_n$ with a hard core repulsion for $n < m$.
$\cale(m)$ gives the
change in free energy from unzipping exactly $m$ bases under the
influence of a force
$F$.  It can be written as the sum of the free energy $2 m g(F)$ of
the $m$ liberated base pairs and of the change in free energy of the
dsDNA when it is shortened by $m$ base pairs.  This second term takes
account of any fluctuations that open base pairs beyond the first closed base $c_1$ and is
independent of $F$.  For homopolymeric DNA, this term takes the form $-m g_0$, where $g_0 < 0$ is the average free energy per base pair of
dsDNA.  Once sequence heterogeneity is present, however,
we must include sequence-dependent deviations from the average.  If the deviation from the average on opening the $n^{\text{th}}$
base is $\eta(n)$, then $\cale(m)$ can be written as
\be
\cale(m) =  [2 g(F) - g_0] m + \sum_{n=1}^m \eta(n) \; .
\ee
Consider now the statistics of the random contribution $\eta(n)$, assuming that the underlying DNA
sequence is random and uncorrelated.  The function $\eta(n)$ reflects this
bare sequence [represented by \ett\ in the continuum model potential~\parenref{v_of_n}] dressed by thermal fluctuations.  As long as the dsDNA is well below its melting
temperature, one expects
that $\eta$ will be a random variable with correlations that
decay on the scale of the finite correlation length of the dsDNA.  If we are only
interested in long length-scale properties, we can thus take $\eta$ to
be Gaussian white noise.  It is convenient to define a quantity 
\be
f \equiv  2 g(F) - g_0 \; ; \label{f-defn}
\ee
$f$ is positive below the unzipping transition and negative above
it.
Passing to the
continuum limit, we can then write
\be
\cale(m)   =   f m + \int_0^m d\!n \eta(n)  \label{cale-defn} \; ,
\ee
where $\eta(n)$ is a zero mean Gaussian random variable which satisfies
\be
\overline{\eta(n) \eta(n') }  =  \Delta \delta(n - n') \; . \label{eta-defn}
\ee
Here the overbar indicates a ``disorder average'' over different
realizations of the random base
sequence.  The associated partition function is simply, up to an
unimportant multiplicative constant,
\be
Z = \int_0^\infty d\!m  \exp\left( -\frac{\cale(m)}{\ktm} \right) \; . \label{Z-defn}
\ee

Eqs. \parenref{cale-defn} through \parenref{Z-defn} define the basic model that
we will study for the remainder of this paper.  It is simple enough to
allow a number of exact predictions, but still correctly captures the
coarse-grained features of unzipping in the presence of sequence
heterogeneity.  It is not difficult
to see that our model shows a sharp unzipping transition:  At
$F=0$, $f = 2 g(0) - g_0 = -g_0$ is positive.  As the pulling force
$F$ is increased from 0, $g(F)$ becomes negative, and $f$ decreases but
remains positive. $\cale(m)$ thus grows
linearly for large $m$, and at most a finite number of bases near the end
of the dsDNA can be unzipped.  These do not contribute appreciably to the average
free energy per base pair of a very long molecule, which remains $g_0$
as at zero force.  As $F$ increases and $g(F)$ becomes increasingly
negative, however, $f$ changes sign
at some critical force value \fc\ satisfying 
\be
2 g(\fc) = g_0 \; . \label{fc-defn}
\ee
Upon expanding about \fc\, we see that to leading order, $f \sim \fc - F$.  For $F>\fc$, the average slope $f$ of $\cale(m)$ is
negative, and $\cale(m)$ tends towards negative infinity for large $m$.
It is thus advantageous to unzip the dsDNA completely.  With all base-pairs unzipped, the average free energy per pair becomes $2
g(F)$.  The discontinuous slope at \fc\ of the free energy per base
pair as a function of $F$ (see Fig.~\ref{first-order}) indicates that the bulk transition is
first order.  Surface quantities
such as $\mtherm$ will nonetheless diverge as the transition is
approached, just as in a critical wetting transition near a conventional
first order phase transition~\cite{wetting-reviews}.
The precise surface behavior in this one-dimensional system will be the subject of subsequent sections.

For dsDNA in physiological conditions, one can ignore the rare
fluctuational openings of base pairs in the bulk and use published
base pairing energies to estimate the parameter values in our model.
The pairing energies typically vary between roughly 1 and 3 \kt\ per
base~\cite{dna-en}; one thus finds $g_0 \sim 2 \kt$ and $\Delta \sim 1
(\kt)^2$.  A typical Kuhn length for single-stranded DNA (ssDNA) is $b \sim 15
\AA$~\cite{busta,bensimon}; inserting this value into the freely
jointed chain expression for $g(F)$ (Eq.~\ref{fjc}) gives a pulling
force $F$ of order $10 \text{pN}$ at the unzipping transition.  As we shall
see, the sequence randomness dominates when $|f|
\lesssim \Delta/\kt \sim \kt$;
randomness is hence important whenever there is appreciable unzipping in
heterogeneous polynucleotide sequences in physiological conditions.

The model of Eqs.~\parenref{cale-defn} through \parenref{Z-defn} is
considerably more general than the semi-microscopic models from which
we derived it.  For example, once $g(F)$ has been ascertained [e.g. by
measuring the force-extension curve of ssDNA in appropriate
conditions~\cite{liphardt,busta,bensimon} and using
Eq.~\parenref{g(F)-gen}], it can be used without reference to any
underlying description of the ssDNA.  In fact, many predictions of our
model are independent of its exact form.  Similarly, most models of
dsDNA (or of RNA hairpins) can be used to define parameters $g_0$ and
$\Delta$; all of the relevant information about the duplex is
contained in these two numbers.  We also expect that our description
applies even when non-local interactions along the ssDNA backbone are
allowed.  All that is required is that the free energy of the ssDNA be
proportional to $m$, so that a function $g(F)$ can be defined.  For
example, a polymer in a good solvent under tension can be described as
a string of blobs~\cite{pincus}.  Once $m$ is larger than the blob
size, as must occur close enough to the unzipping transition, the free
energy of the single strands will be proportional to $m$.  In fact, in
physiological conditions and at the forces of order 10 pN required to
unzip dsDNA, the blob size will be at most a few monomers, meaning
that excluded volume interactions can be neglected entirely in a first
approximation.  Likewise, a model of single stranded polynucleotides
with uniformly attractive, non-random base pairing interactions
(tending to produce hairpins) predicts a free energy proportional to
the number of bases in the strand~\cite{mezard}.  This model agrees
well with experimental force-extension curves for ssDNA.  The same
calculations show that the fraction of bases in the liberated single
strands involved in intra-strand pairing interactions will be small at
the unzipping transition.  Sequence variation will further suppress
such pairing: Because not all bases can pair with each other, it will
generally be necessary to make a large loop in order to bring together
two stretches of bases that can pair to form a stem.  This means that
more work must be done against the pulling force for the same gain in
base pairing energy.  Although it might still
be possible for a stem region of atypically high GC content to pair in
this way, in a truly random sequence the probability of finding such a
region decays exponentially with its length.

\subsection{Related physical systems}


Although the main focus of this paper will be the mechanical unzipping
of polynucleotide duplexes, our formalism also applies to
other experiments and physical systems.  For example, an alternative
method for unzipping DNA is to force
one of the single strands through a very small pore by
applying an electric field~\cite{dan-personal}.  If the pore is so
narrow that double-stranded DNA cannot fit through it, and if the
applied field is strong enough, one of the single strands can enter the
pore and be drawn through it, thereby unzipping the duplex (see
Fig.~\ref{pore-unzip}).  In this case, the analog of $g(F)$ is the
electrostatic energy gained by the single strand passing through the
pore, reduced by any entropic penalty the other single strand
must pay due to confinement by parts of the pore or the adjoining
walls~\cite{me-biophysj}.  Continuum models such as Eq.~\parenref{hc} are also commonly used
to describe a number of other systems; in several of them, there is a
natural analog to the pulling force \fc.  Examples include the
adsorption of a Gaussian random heteropolymer, where \fc\ maps directly
to a force pulling the end of the polymer away from the adsorbing
surface~\cite{jf}, and a flux line in a type II superconductor bound to a
fragmented columnar defect~\cite{denis}, where \fc\ can be viewed as the magnetic
field strength perpendicular to the defect.  In
addition, the Hamiltonian ${\cal H}_{\rm C} + {\cal H}_{\rm C,pull}$
bears a strong resemblance to models of the wetting transition in two
dimensions in a wedge with angle close to 180 degrees~\cite{wetting}.

\section{Statistical Mechanics of Homopolymer Unzipping}
\label{non-rand-sect}

Before tackling the more difficult problem of unzipping a
double-stranded molecule with a random base sequence, we describe some results for a uniform sequence~\cite{me-prl}.  If the energy
cost of opening each successive base pair is the same, then the
deviation $\eta(n)$ from the average vanishes identically, and
$\cale(m) = f m$.  Even if, as would be the case for an alternating
base sequence, $\eta(n)$ is a non-zero periodic function, we expect
that on scales longer than its period, $\eta(n)$ can safely be set to
zero.  In this section, we show explicitly that the semi-microscopic
continuum model discussed above [Eqs.~\parenref{hc}
and~\parenref{hcpull}] gives results identical to those following from
the simpler single degree of freedom description.

Equilibrium statistical mechanics in a linear potential is
straightforward.  The partition function of our minimal model is simply $Z = \int_0^\infty
d\!m \exp(-m f/\kt) = \kt/f$, and the probability of opening exactly $m$
bases is $(f/\kt) \exp(-m f/\kt)$.  The equilibrium moments of $m$ can be obtained from
derivatives of the free energy $G(f) = -\kt \ln Z$ with respect to $f$:
$\langle m \rangle = \partial G/\partial f= \kt/f$, $\langle m^2 \rangle - \langle m
\rangle^2 = \partial^2 G/\partial f^2 = \kt/f^2$, and so on.  Recalling that $f \sim \fc - F$, we
see that \mavg\ exhibits a power law divergence near the unzipping transition;
\be
\langle m \rangle \sim (\fc - F)^{-1} \; \; \; \;
\text{(homopolymer).} \label{m-homopoly}
\ee
The divergence of $\langle m \rangle$ has a simple origin:  Although the absolute minimum of $\cale(m)$ remains at $m
= 0$ everywhere below the transition, the system explores all
configurations with $\cale(m) \lesssim \kt$, or equivalently $ m
\lesssim \kt/f$; this of course suggests the same scaling for \mtherm\
found in the exact
calculation. The homopolymer thus opens partially for $F < \fc$
entirely in
order to gain {\em entropy}.  We shall see in subsequent sections that
a very different physical mechanism dominates in the unzipping of heteropolymers.

\subsection{Connection to non-Hermitian Delocalization}
\label{nhqm-sectn}

A different perspective on the mechanical denaturation of a
homopolymer follows from viewing the energy $\hc + \hcp$ of the
continuum model [Eqs.~\parenref{hc} and~\parenref{hcpull}] as an imaginary time quantum mechanical action.
The partition function $Z(\rlb,N)$ of a strand of
length $N$, subject to the constraint $\rlb(N) = \rlb$, satisfies the
partial differential equation~\cite{doi-ed}
\be
\frac{\partial Z}{\partial N} =
\frac{b^2}{d}\left(\nabla_{\rlb}+\frac{\fb}{\ktm} \right)^2 Z - \frac{V(\rlb)}{\ktm} Z
\equiv - {\cal L}(\fb) Z \; , \label{nhqm-schrod}
\ee
where the sequence-dependent function $V_N(\rlb)$ is replaced by the
$N$-independent potential
$V(\rlb)$ for a homopolymer.    In order to avoid a proliferation of factors of
$a/b$, we assume that the backbone distance $a$ between chemical
monomers is equal to the Kuhn length $b$.
When $\fb = \bbox{0}$, Eq.~\parenref{nhqm-schrod} is just an imaginary-time
Schr\"{o}dinger equation.  With the addition of a nonzero pulling
force \fb, the strict correspondence with conventional quantum mechanics is lost.
Nonetheless, much can be learned by studying the
evolution operator ${\cal L}$ using the language of quantum mechanics.  This avenue as been pursued for the formally identical
problem of a flux line pinned to a defect in a type II
superconductor~\cite{hatano-nelson}.  In this subsection, we show
explicitly that results from this more microscopic approach can be
recovered from the simplified model embodied in Eqs.~\parenref{cale-defn}
through~\parenref{Z-defn}.

In analyzing Eq.~\parenref{nhqm-schrod}, it is useful to view the force
\fb\ as a constant, imaginary vector potential.  The ``gauge
transformation'' operator
\be
{\cal U} : \; \psi(\rlb) \mapsto \exp(\fb \cdot \rlb/\kt) \psi(\rlb)
\ee
 can thus be
used to relate the operator ${\cal L}(\fb)$ at a force \fb\ to the
Hermitian operator ${\cal L}(\bbox{0})$:
\bea
{\cal L}(\fb) & = & {\cal U} {\cal L}(\bbox{0}) {\cal U}^{-1} \; ;
\nonumber \\
\call(\bbox{0}) & = & -\frac{b^2}{d} \nabla_{\rlb}^2 + \frac{V(\rlb)}{\ktm} \; .
\eea
Under the same transformation, the eigenfunctions $\psi_n^{\bbox{\rm F}}(\rlb)$ of
${\cal L}(\fb)$ are given by
\be
\psi^{\bbox{\rm F}}_n(\rlb) = {\cal U} \psi^{\bbox{0}}_n(\rlb) =
e^{\bbox{\rm F} \cdot \bbox{\rm r}/\ktm} \psi^{\bbox{0}}_n(\rlb) \; . \label{efunct-transf}
\ee
Eq.~\parenref{efunct-transf} shows that
exerting a non-zero force \fb\ biases the eigenfunctions in the
direction of the force.
This transformation is valid as long as the new eigenfunction
$\psi^{\bbox{\rm F}}_0$ satisfies the same boundary conditions as the
untransformed eigenfunction.  If we think of an isolated polymer in a
box whose size tends towards infinity, the appropriate boundary conditions are that
$\psi^{\bbox{\rm F}}_n$ be well behaved at infinity; given the form of ${\cal U}$,
this is equivalent to demanding that the eigenfunction
$\psi_n^{\bbox{0}}(\rlb)$ of the Hermitian problem decay at
least at as fast as $\exp(-F R/\ktm)$ for large $R = |\rlb|$.  When this
condition holds for the $n^{\text{th}}$ eigenfunction, the
corresponding eigenvalues of $\call(\bbox{0})$ and $\call(\fb)$
will be identical, and the eigenfunctions will be related according to
Eq.~\parenref{efunct-transf}.  Because, according to Eq.~\parenref{nhqm-schrod} the contribution of each eigenvalue
$\lambda_n$ to the partition function decays like $\exp(-\lambda_n
N)$, the smallest eigenvalue
$\lambda_0$ dominates in the limit of a very long polymer duplex.  We are interested in conditions in which the dsDNA is stable in
the absence of a pulling force; in this case, $\call(\bbox{0})$, which
describes the native, unpulled polymer, must have at least one bound state.
The ground state eigenvalue $\lambda_0 < 0$ differs from the free energy
per length $g_0$ of dsDNA introduced previously only by a factor of
\kt : $g_0 = \kt \lambda_0$.  Because $V(\rlb)$ is a
short-ranged potential, the ground state wavefunction $\psi_0^{\bbox{0}}(\rlb)$
should decay like $\exp(- \kappa_0 R)$ for large $R$, with the decay
rate given by
\be
\kappa_0 = \frac{1}{b} \sqrt{|\lambda_0| d} = \frac{1}{b} \sqrt{\frac{|g_0| d}{\ktm} } \; ,
\ee
where $d$ is the spatial dimension.
When applied to the ground state wavefunction, the gauge transformation of Eq.~\parenref{efunct-transf} thus breaks down
at a force of magnitude \fc\ given by
\be
\frac{\fc}{\ktm} = \kappa_0 \Leftrightarrow \fc  = \frac{\ktm}{b}
\sqrt{\frac{|g_0| d}{\ktm} } \; . \label{nhqm-fc}
\ee

It is natural to regard this force as the location of the unzipping
transition.  Indeed, one can show~\cite{hatano-nelson} that far from
the ends of a long polymer, the probability that a given base pair
will be separated by a displacement $\rlb$ is $P_\infty(\rlb) =
\psi_0^{\bbox{\rm F}}(\rlb) \psi_0^{\bbox{-F}}(\rlb)$. For $F < \fc$, the two gauge transformations cancel each
other, and $P_\infty(\rlb) = [\psi_0^{\bbox{0}}(\rlb)]^2$.  Thus, below
\fc\, paired bases in the bulk of the dsDNA always stay near each
other, and the polymer is below the unzipping transition (see
Fig.~\ref{radial-distr}).  Conversely, above \fc, where the gauge transformation
is no longer valid, the eigenfunctions $\psi_0^{\pm \bbox{\rm F}}$ are
dominated by \fb\ and are extended.  Indeed, one can demonstrate that they become plane waves as $R \rightarrow \infty$.  The two single strands are then typically
widely separated (Fig.~\ref{radial-distr}), and the DNA is above an unzipping transition given
by Eq.~\parenref{nhqm-fc}. 

Upon inserting the expression for $g(F)$
[Eq.~\parenref{g-G}] appropriate for the Gaussian single-stranded
polymer into our previous
criterion $2 g(\fc) = g_0$, we obtain a value for the critical
unzipping force \fc\ identical to Eq.~\ref{nhqm-fc}.  In fact, provided the duplex binding potential
$V(\rlb)$ vanishes as $R \rightarrow \infty$, $\psi_0^{\bbox{\rm F}}(\rlb)$ will
approach a nonzero constant for large $R$ above \fc.  One can then read off
$\lambda_0 =  -b^2 F^2/d (\kt)^2$ directly from
Eq.~\parenref{nhqm-schrod}; the free energy above the transition is
simply $\kt \lambda_0 = 2 g(F)$, a natural result given that above
the unzipping transition the DNA is entirely in the single-stranded
form.  Within the present formalism, one can also obtain a closed-form
expression for $\lambda_0$, and hence for the free energy per monomer,
below the unzipping transition.  For $F < \fc$, the
transformation~\parenref{efunct-transf} is valid, and $\lambda_0 =
-b^2 \kappa_0^2(T)/d$, independent of $F$.  Both the entropy, given by a derivative of
$\kt \lambda_0$ with respect to $T$, and the average extension per
nucleotide in the bulk, given by a derivative of $\kt \lambda_0$ with
respect to $F$, change discontinuously at \fc\ (see Fig.~\ref{first-order}).
The bulk unzipping transition is thus
first order, as is the case for the related problem of a single flux
line torn away from a columnar defect in a type II superconductor~\cite{hatano-nelson}.

Because $\kt \lambda_0$ is the {\em bulk} free energy per monomer, its
derivatives tell us nothing about the diverging surface precursors to
the unzipping transition.  To study surface effects within the quantum
mechanical formalism, note that the probability that the
ends of the two single strands are separated by a displacement $\rb = \rlb(0)$ is
given by~\cite{hatano-nelson} 
\be 
P_0(\rb) = \psi_0^{\bbox{\rm F}}(\rb )
\simeq \exp\left[ \frac{\fb \cdot \rb}{\ktm} - \kappa_0 | \rb |
\right] \; , \label{nhqm-rdistrn} 
\ee 
where the last equality is valid
outside the range of the potential $V(\rlb)$.  Focusing, for simplicity,
on the case of one spatial dimension ($d=1$), and replacing the
vectors \rlb\ and \rb\ by the scalars $R$ and $r$, it follows that the
average distance $\langle r \rangle$ between the ends of
the two single strands diverges like 
\be 
\langle r \rangle = \frac{
(\kappa_0 - F/\ktm)^{-2} + (\kappa_0 + F/\ktm)^{-2}}{(\kappa_0 -
F/\ktm)^{-1} + (\kappa_0 + F/\ktm)^{-1} } \sim \frac{1}{\fc - F}\; .
\ee 
Slightly more involved calculations~\cite{hatano-nelson} give
the decay of the end to end distance as the bulk value is approached: 
\be
\langle R(n) \rangle = \langle r \rangle \exp\left( -\frac{n}{n^*}
\right) \; , \label{avg-rn} 
\ee 
where $\langle R(n) \rangle$ is the average distance between the two single
strands at base pair $n$.  The healing length $n^*$ diverges near \fc\ as
\be
 n^* = \frac{\ktm^2}{b^2 (\fc^2 - F^2)} \sim \frac{1}{\fc - F}\;
. \label{nstar} 
\ee

To check these results against the single degree of freedom model defined by Eqs.~\parenref{cale-defn}--\parenref{Z-defn}, one must
translate the number of unzipped base pairs $m$ into a distance $r$ between the ends of the two
single strands.  When $m$ base pairs have been
unzipped, $r$ is simply the end to end distance of a Gaussian
polymer of length $2 m$ subject to a force $F$; it thus has
distribution~\cite{pincus}
\be
P_0(r|m) = \frac{1}{\sqrt{4 \pi m b^2} } \exp\left\{ -\frac{[r - 2
m b^2 F/\ktm]^2}{4 m b^2} \right\} \; . \label{pzero-rm}
\ee
The probability that precisely $m$ base pairs have been unzipped is
$P(m) = (f/\kt) \exp(-m f/\kt)$, so the full distribution of $r$ is
given by
\be
P_0(r) = \int_0^\infty d\!m \, P(m) P_0(r | m) \; . \label{pzero-r}
\ee
Evaluating this integral leads to the prediction summarized in
Eq.~\parenref{nhqm-rdistrn}.  Similarly, the distribution $P_n(R)$ of $R(n)$ for any
$n$ can be obtained by summing over a conditional distribution, assuming that
$m>n$ bases are open, and another one given that $m<n$ bases are open.  The latter
distribution is well-approximated, except for $n$ very near $m$, by the
bulk distribution for dsDNA $P_\infty[R(n)]$ introduced earlier.
Thus, we find that
\bea
P_n(R) & = & \int_n^\infty d\!m \, P(m) P_0(R | m-n) + P_\infty(R)
\int_0^n d\!m \, P(m)
 \nonumber \\
& = & \exp\left( -\frac{n f}{\ktm} \right) P_0(R) + \frac{f}{\ktm} \left[ 1 -
\exp\left( -\frac{n f}{\ktm} \right) \right] P_\infty(R) \; , \label{pn-r}
\eea
where $P_0(R|m-n)$ and $P_0(R)$ are given by Eqs.~\parenref{pzero-rm} and
\parenref{pzero-r}.  Since $P_\infty(R)$ must be symmetric with respect to
$R=0$, its average vanishes.  Upon using Eq.~\parenref{pn-r} to
evaluate $\langle R(n) \rangle$
and recalling that $f = |g_0| - 2 g(F) = (\fc^2 - F^2) b^2/(\kt)^2$ for
a Gaussian chain, we recover Eqs.~\parenref{avg-rn} and
\parenref{nstar}.  Thus, the predictions obtained by
studying directly the evolution equation~\parenref{nhqm-schrod} of the
partition function coincide with those obtained by integrating out most degrees of freedom to
arrive at a simplified formulation in terms of the unzipping energy $\cale(m)$.

\section{Disorder-Averaged Behavior}
\label{disorder-average}

In contrast to the
entropically-driven opening of a homopolymer, the unzipping of a
polymer with a random sequence is driven primarily by the possibility
of lowering $\cale(m)$ by unzipping a string of base pairs that are
more weakly paired than the average.  The two
transitions are thus qualitatively different. To see this explicitly, consider a simple application of the Harris
criterion for the importance of disorder~\cite{harris}.  The typical variation per monomer due to
disorder in the
base-pairing energy $\cale(m)$ of a liberated section of length $\langle m
\rangle$ is $(\Delta/\langle m \rangle)^{1/2} \sim \sqrt{\fc - F}$, where the
$F$ dependence follows from the result~\parenref{m-homopoly} for the
divergence of $\langle m \rangle$ near the transition for a
homopolymer.  These energy variations vanish more slowly as $F
\rightarrow \fc$ than the average energy difference $f \sim \fc - F$
between the two phases, indicating that sequence randomness dominates at the unzipping transition.  

A related argument can help us
to guess the correct critical exponent for the divergence of \mtherm\
when disorder is present: The contribution to $\cale(m)$
of the average energy difference is $m f$, while a typical
favorable contribution from random variations about the average is of order
$-\sqrt{\Delta m}$.  The random part thus exceeds the average for $m
\lesssim m^* \equiv \Delta/f^2$.  When this is the case, $\cale(m)$ is
roughly as likely to be negative as to be positive.  One thus expects
that a typical value of $\langle m \rangle$ will be at least of order
$m^*$.  Near enough to the unzipping transition at $f=0$, $m^*$ is larger than
the equilibrium average $\kt /f$ for a non-random sequence.  Instead of the $1/(\fc - F)$ divergence in \mtherm\
seen for a homopolymer, one might thus expect DNA with a random sequence to show a
considerably stronger $1/(\fc - F)^2$ singularity.  The crossover between
the two scaling regimes should occur when $\Delta/f^2 \sim \kt/f$, or
when $f \sim \Delta/\kt$.  For dsDNA, both $\sqrt{\Delta}$ and the
average base pairing energy $g_0$ are of order \kt; we can estimate
$f \approx g'(\fc) (\fc - F) \approx (g_0/\fc) (\fc - F)$.  Hence,
when $f \kt /\Delta \sim {\cal O}(1)$ at the crossover, the reduced force
$(\fc - F)/\fc$ is also ${\cal O}(1)$, confirming that disorder cannot be
neglected in polynucleotide unzipping even for $F$ of order, say,
$\fc/2$.  As we shall see in Section~\ref{dynamics}, disorder affects
the {\em dynamics} of unzipping for a similar range above \fc.

This scaling argument can be extended to the case of random DNA sequences
with long-ranged correlations (as may be the case for noncoding
DNA~\cite{dna-stat}).  If the correlations between nucleotides
separated by $m$ base pairs decay like $1/m^{\gamma}$,
the fluctuation $\eta(m)$ around the average energy to open a base pair
will likewise have a correlation function $\overline{\eta(m) \eta(m')}
\sim 1/|m-m'|^{\gamma}$.  For $\gamma < 1$, the mean-squared value of
$\int_0^m d\!m' \eta(m')$ then grows like $\int_0^m d\!m' \int_0^m
d\!m'' 1/|m'-m''|^{\gamma} \sim m^{2-\gamma}$.  A typical random
contribution to $\cale(m)$ then increases as $m^{1-\gamma/2}$; balancing
this random energy against $m f$ suggests that $\mavg \sim m^* \sim f^{-2/\gamma}$.
If we take, for example, $\gamma = 2/3$~\cite{dna-stat}, then $\mavg
\sim 1/f^3$, an even stronger divergence.

To verify our scaling argument for the case of a random, uncorrelated
base sequence, we begin by calculating the
disorder-averaged number of bases opened \mavg\ (as before, the
overbar indicates an average over different random base sequences).
Fluctuations about this average will be studied in more detail in the
next section.  To find \mavg, one must first compute the average free
energy $-\kt \overline{\ln Z}$; disorder-averaged cumulants of $m$ can
then be obtained by taking derivatives with respect to $f$.  Remarkably, the entire distribution of $Z$ can be found exactly by treating
the random energy $\eta$ as a Langevin noise.  Several variations on
this procedure have appeared in other
physical contexts~\cite{rand-fp}, as have related approaches to the
same formal problem~\cite{rand-moments}.  

We begin by
defining the partition function of a polymer of finite length $m$:
\be
\zt(m) = \int_0^m d\!m' \exp\left[ -\frac{\cale(m')}{\ktm} \right] \;
.
\ee
The partition function $Z$ of interest to us is recovered by taking
the limit of an infinite length polymer: $Z = \lim_{m \rightarrow
\infty} \zt(m)$.  The derivative of \zt\ is simply
\be
\frac{d \ztm}{dm} = e^{-\cale(m)/\ktm} \; , \label{zt-ode}
\ee
with initial condition $\zt(0) = 0$.  Similarly, the derivative of
$\cale(m)$ is, from Eq.~\parenref{cale-defn},
\be
\frac{d \cale}{dm} = f + \eta(m) \;, \label{cale-ode}
\ee
with initial condition $\cale(0) = 0$.  Eqs.~\parenref{zt-ode} and
\parenref{cale-ode} make up a system of coupled Langevin
equations, analogous, for example, to those describing the Brownian
motion of a massive particle, with $\cale$ playing the role of
momentum and \zt\ that of position.  They can be transformed in the
usual manner into an
equivalent Fokker-Planck equation for the joint probability distribution
$P(\cale,\zt,m)$ of $\cale$ and \zt\ at ``time'' $m$~\cite{vank}:
\be
\frac{\partial P}{\partial m} = \left[ \frac{\Delta}{2}
\frac{\partial^2}{\partial \cale^2} - f \frac{\partial}{\partial \cale} - e^{-{\cal E}/\ktm}
\frac{\partial }{\partial \zt} \right] P \; .
\label{fp-eqn}
\ee

To solve Eq.~\parenref{fp-eqn} in the limit of large $m$, we first Laplace transform with respect to \zt\ and to $m$, with
conjugate variables $\lambda$ and $s$, respectively.  The resulting
ordinary differential equation for the transformed distribution
$\hat{P}(\cale;\lambda,s)$ takes the form
\be
\frac{\Delta}{2} \frac{d^2 \phat}{d \cale^2} -f
\frac{d \phat}{d \cale} - \lambda e^{-\cale/\ktm} \phat
- s \phat = - \delta(\cale) \; .
\ee
The change of variables
\be
x \equiv \kt \left( \frac{8 \lambda}{\Delta} \right)^{1/2} e^{-\cale/(2
\ktm)} \;
\ee
leads to an inhomogeneous Bessel equation
\be
x^2 \frac{\partial^2 \phat}{\partial x^2} + \left(1 + \frac{4
f \ktm}{\Delta} \right) x \frac{\partial \phat}{\partial x} -  \left[ x^2 +
\frac{8 s (\ktm)^2}{\Delta} \right] P = -\frac{4 x_0 \ktm}{\Delta} \delta(x - x_0) 
\; , \label{ode-x}
\ee
where $x_0 \equiv x|_{\cale = 0} = \kt \sqrt{8 \lambda/\Delta}$. Although $\cale$ has been replaced by $x$, $\phat$ remains normalized as
a function of $\cale$.  One can
easily check that the solution of Eq.~\parenref{ode-x} follows the usual
form for the Green's function of a Sturm-Liouville equation:
\be
\phat(x; \lambda, s) =\kt \left\{
\begin{array}{ll} 
\frac{4}{\Delta} \left( \frac{x_0}{x} \right)^{2f \ktm/\Delta} K_{\nu}(x_0)
I_{\nu}(x), & x \leq x_0 \\
\frac{4}{\Delta} \left( \frac{x_0}{x} \right)^{2f \ktm/\Delta} I_{\nu}(x_0)
K_{\nu}(x), & x \geq x_0 \; ,
\end{array} 
\right.
\label{soln-x}
\ee
where $I_\nu$ and $K_\nu$ are modified Bessel functions, and
\be
\nu = \kt \sqrt{\frac{8 s}{\Delta} + \frac{4 f^2}{\Delta^2} } \; .
\ee

Eq.~\parenref{soln-x} represents an exact solution to our
single degree of freedom model.  We are interested primarily in the distribution
of \zt\ for large $m$, so we would like to integrate over all $\cale$
and then take the limit $m \rightarrow \infty$.  The first task can
easily be accomplished on a formal level:
\bea
\phat(\lambda, s) & = & \int_{-\infty}^{\infty} d\!\cale \phat(\cale,\lambda; s)
\nonumber \\
& = & \frac{8 (\ktm)^2}{\Delta} \left[ K_{\nu}(x_0) \int_0^{x_0} \frac{d\!x}{x}
\left(\frac{x_0}{x}\right)^{2 f \ktm/\Delta} I_{\nu}(x) + I_{\nu}(x_0)
\int_{x_0}^{\infty} \frac{d\!x}{x} \left(\frac{x_0}{x}\right)^{2
f \ktm/\Delta} K_{\nu}(x) \right] \; . \label{p-int}
\eea
Because $\cale(m)$ grows linearly with $m$ below the unzipping
transition for large enough $m$,
the contributions to the partition function $Z$ of the parts of the
dsDNA at very large $m$ are exponentially suppressed.  Hence, we
expect that \zt\ must have a well-defined limiting distribution as $m
\rightarrow \infty$.  This, in turn, implies that the Laplace
transform $\phat(\lambda, s)$
should diverge like $1/s$ as $s \rightarrow 0$, or equivalently, as
$\nu \rightarrow 2 f \ktm/\Delta$.  An examination of Eq.~\parenref{p-int}
reveals that this is in fact the case.  Specifically, $I_\nu(x) \sim
x^{\nu}$ for small $x$, so the integral from $0$ to $x_0$ diverges when
$\nu$ approaches $2 f \ktm/\Delta$.  This singularity dominates the large
$m$ behavior of the inverse Laplace transform with respect to $s$,
allowing us to perform the inversion analytically:
\be
\phat(\lambda; m \rightarrow \infty) = \frac{2}{\Gamma(2f \ktm/\Delta)}
\left[\frac{2 \lambda (\ktm)^2 }{\Delta}\right]^{f \ktm/\Delta} K_{2
f \ktm/\Delta}\left(\ktm \sqrt{\frac{8 \lambda}{\Delta} } \right) \; , \label{p-largem}
\ee
where we have substituted $x_0 = \kt \sqrt{8
\lambda/\Delta}$.  Note that the asymptotics are
completely determined by the small $x$ behavior of
$\phat(x;\lambda,s)$.  Because small $x$ corresponds to large $\cale$,
this is quite reasonable:  It follows directly from
Eq.~\parenref{cale-ode} that the distribution of $\cale(m)$ is
a Gaussian centered at $m f$, so only very large $\cale$ will have any weight for
large $m$.

To evaluate the disorder-averaged free energy, we must invert the Laplace transform $\phat(\lambda;m
\rightarrow \infty)$ to obtain the distribution $P(Z)$ of the partition
function. With the aid of various Bessel function
identities, one discovers that the integral can be evaluated
analytically.  The result is the distribution over possible random
sequences of the partition
function $Z$ of our minimal unzipping model~\cite{rand-fp}:
\be
P(Z) = \frac{1}{\Gamma(2f \ktm/\Delta)} \left[ \frac{2 (\ktm)^2}{\Delta}
\right]^{2f \ktm/\Delta} \left( \frac{1}{Z} \right)^{1+2f \ktm/\Delta}
\exp\left[\frac{-2 (\ktm)^2}{Z \Delta}\right] \;.
\ee
The disorder-averaged free energy follows immediately by integration;
with the substitution $y \equiv 2 (\kt)^2/(Z \Delta)$, one has
\be
-\kt \overline{\ln Z} = \kt\left\{ \frac{1}{\Gamma(2 f \ktm/\Delta)}
\int_0^{\infty}
d\!y \, y^{2f \ktm/\Delta - 1} \ln(y) e^{-y}  + 
\ln\left[\frac{\Delta}{2 (\ktm)^2}
\right] \right\}  \; .
\ee
Taking a derivative with respect to $f$  yields the main quantity of
interest:
\bea
\overline{\langle m \rangle} & = & - \kt \frac{\partial \overline{\ln
Z}}{\partial f} \nonumber \\
& = & \frac{2 (\ktm)^2}{\Gamma(2 f \ktm/\Delta) \Delta} \int_0^{\infty} d\!y \,
y^{2f \ktm/\Delta -1} (\ln y)^2 e^{-y} - \frac{2 (\ktm)^2 \Gamma'(2
f \ktm/\Delta)^2}{\Gamma(2 f \ktm/\Delta)^2 \Delta} \; , \label{mavg}
\eea
where $\Gamma'(z) = d\Gamma/dz$.  This function is plotted
in Figure~\parenref{crossover}.  In agreement with our earlier scaling argument,
there is a crossover from $1/f$ to $1/f^2$ behavior at $f$ of order
$\Delta/\kt$.  Indeed, one can analytically extract the asymptotic
small $f$ behavior from Eq.~\parenref{mavg}.  One finds that to leading
order as $f \rightarrow 0$,
\be
\mavg \simeq \frac{\Delta}{2 f^2} \sim \frac{1}{(\fc - F)^2} \; \; \;
\text{(random heteropolymer).} \label{mavg-asympt}
\ee
Additional results follow for the higher cumulants of
$m$.  For example, the disorder-averaged variance of $m$ can be found
from the second derivative of $\overline{\ln Z}$.  For small $f$,
$\overline{\langle m^2 \rangle - \langle m \rangle^2} = \kt \partial
\overline{\ln Z}^2/\partial f^2 \sim 1/f^3$.  The square root of this
variance is a length scale that can be compared to \mavg.  In the
non-random case, both quantities are of order $\kt/f$.  In contrast, once sequence
randomness is added, we have that $( \overline{\langle m^2 \rangle -
\langle m \rangle^2} )^{1/2} \sim 1/f^{3/2}$, which is much smaller
than \mavg\ for sufficiently small $f$.  Thermal fluctuations about $\langle
m \rangle$ in a given random heteropolymer thus become small compared
to the mean near the transition.  As we shall see in the next section,
this fact allows us to predict not just disorder-averaged quantities (it
might be tedious to average over all possible sequences in a real experiment!),
but also the unzipping behavior of a {\em single} dsDNA molecule.

\section{Force-Displacement Curve for a Single Polynucleotide Duplex}
\label{single-mol}

Figure~\ref{four-polys} plots the average number of unzipped bases $\langle
m \rangle$ versus force near the unzipping transition for simulations
of four different dsDNA molecules, with different
random sequences~\cite{sim-note}.  The corresponding energy landscapes for a force close to \fc\ are shown in
Fig.~\ref{four-ens}.  Far from being smooth,
each \mtherm\ versus $f$ curve shows long plateaus,
where \mtherm\ remains essentially constant, separated by sudden, large
jumps.  The smoothly diverging precursor to the phase transition seen in
homopolymers and in the disorder-average \mavg\ has evidently been replaced by a series of
``micro-first-order transitions.''  The four traces, moreover, are not the same---the unzipping of
a single random dsDNA does not exhibit self-averaging, but instead
shows large sequence-dependent variations.  Most equilibrium systems with
quenched disorder are self-averaging because the macroscopic observables of
interest are the sums of contributions from many essentially
independent correlation volumes, each with their own independent
realization of the quenched random variables; the central limit
theorem then guarantees that in the thermodynamic limit, measurements
will always coincide with the disorder average.  In a single molecule
DNA unzipping experiment, in contrast, one is probing only one
realization of the quenched random sequence.  As Figure~\ref{four-ens}
indicates, each random realization of $\cale(m)$ will be different,
and the value of \mtherm\ at a given $f$ can thus be expected to
differ from one polymer to the next.  Furthermore, for each sequence,
$\cale(m)$ varies over many tens of \kt; one thus might expect that $m$
would not fluctuate very far from the minima.  Figure~\ref{four-polys}
bears out this idea:  The location \mmin\ of the absolute
minimum of $\cale(m)$ for each value of $f$ coincides remarkably well
with \mtherm.  Because $\cale(m)$ is
usually negative at these minima, the dsDNA gains energy by unzipping
some bases at its end, even below the bulk unzipping transition.  This
mechanism contrasts with the essentially entropic impetus for
surface opening in the case of a homopolymer.  We show in this 
section that, near enough to the transition, \mtherm\ for a given
DNA or RNA duplex coincides with \mmin\ with arbitrary precision and that this
fact can be used to gain a quantitative understanding of the abrupt
jumps seen in Figure~\ref{four-polys}.  We will usually work in the
continuum approximation, with the probability
$P(\cale,m)$ of finding an energy $\cale$ after opening $m$ bases
satisfying a diffusion-like equation,
\be
\frac{\partial P}{\partial m} = \frac{\Delta}{2} \frac{\partial^2
P}{\partial \cale^2} - f \frac{\partial P}{\partial \cale} \; . \label{fp-cale}
\ee
This result follows directly from Eq.~\parenref{cale-ode} or from integrating the full Fokker-Planck
equation~\parenref{fp-eqn} with respect to \zt.  At each $m$, $\cale(m)$
thus has a Gaussian distribution; because our results do not depend
on the tails of this distribution, they should be equally valid for
more realistic, discrete models of dsDNA.

\subsection{Dominance of the Absolute Free Energy Minimum}
\label{zero-temp-fp}

We begin by arguing that, close to the transition, the location \mmin\ of the absolute minimum of
$\cale(m)$ is in fact the same as \mtherm.  More precisely, we wish to
show that, for a random DNA sequence, 
\be
\lim_{f \rightarrow 0} \frac{\mtherm}{\mmin} = 1 \;\; \text{with
probability 1.} \label{avg-min}
\ee
In qualitative terms, one might expect this result to hold because the
scale of $\cale(m)$ grows like the square root of the distance from
the minimum; it is thus very unlikely that $\cale(m)$ will revisit the
neighborhood of its minimum value for $m$ far from the
location of the original minimum.  Here, we simply outline the arguments necessary to support
this intuition; closely related theorems have, however, been proven
with mathematical rigor~\cite{yor}.  We will proceed by first
considering scenarios in which Eq.~\parenref{avg-min} would not hold, then
showing that the probability of each such event vanishes as $f
\rightarrow 0$.  In
renormalization-group language, Eq.~\parenref{avg-min} can be read as
stating that the unzipping transition for a random dsDNA sequence is
governed by a zero-temperature fixed-point; such fixed points have
been found in a number of other random systems~\cite{dsf-terry}.

The simplest way that \mmin\ and \mtherm\ could differ is for \mmin\
to equal 0; since \mtherm\ is necessarily positive, their
ratio would then be infinite.  The probability that $\mmin = 0$ is the same as the
probability that the biased random walk $\cale(m)$, which starts at
$\cale(0)=0$, has $\cale(m) > 0$ for all $m > 0$.  More generally, the
probability that $\cale(m) > 0$ for all $m>0$ for a random walk starting at $\cale(0) = \cale_0$ is known in the literature on first
passage problems as the ``splitting probability''
$\pi(\cale_0)$.   The splitting probability satisfies an equation involving
the adjoint of the diffusion operator~\cite{vank},
\be
\frac{\Delta}{2} \frac{\partial^2 \pi}{\partial \cale_0^2} + f
\frac{\partial \pi}{\partial \cale_0} = 0 \; .
\ee
The solution of this equation with boundary conditions $\pi(0) = 0$ and
$\pi(\infty) = 1$ is $\pi(\cale_0) = 1-\exp( - 2 \cale_0
f/\Delta)$.  The requirement that $\pi(0) = 0$ is an artifact of the
behavior of a continuous time random walk as $m \rightarrow 0$:
Because $\cale(m)$ experiences small jumps up and down on all scales, a
random walk that starts at $\cale(0) = 0$ will pass below the line
$\cale = 0$ many times for very small $m$.  This behavior is not
relevant to real DNA with discrete bases, and we can regularize it by
considering, instead of a random walk that starts exactly at $\cale_0
= 0$, one that starts slightly above 0.  For small $\cale_0$,
$\pi(\cale_0) \approx 2 \cale_0 f/\Delta$, so the splitting probability vanishes linearly as $f \rightarrow 0$.  Indeed, for any $\cale_0$,
$\pi(\cale_0)$ goes to zero linearly for small $f$, as one might expect
based on the well-known result that a completely unbiased random
walk in one dimension must eventually visit the entire real line.  The same linear
behavior for small enough bias is
seen in random walks on one-dimensional lattices~\cite{vank}.  We
thus conclude that the probability that $\mmin = 0$ is proportional to
$f$ and can be neglected as $f \rightarrow 0$.

Now consider other possible values of \mmin. We shall see in the next subsection that the distribution of \mmin\ for
$\mmin>0$ is a function of the dimensionless ratio $\mmin f^2/\Delta$.  The probability
that $\mmin \sim {\cal O}(1/f^{\beta})$, with $\beta \neq 2$, hence
becomes negligible for small $f$, and we need only consider
$\mmin \sim {\cal O}(1/f^2)$.  For the absolute minimum and the thermal
average {\em not} to coincide in this case, there must be a local minimum
nearly degenerate with $\cale(\mmin)$ a distance ${\cal O}(1/f^2)$ away
from \mmin.  Note in particular that a degenerate minimum closer to
\mmin\ than ${\cal O}(1/f^2)$ will contribute an additive
correction to \mtherm\ that is much smaller than $\mmin \sim {\cal
O}(1/f^2)$ for small enough $f$, and thus will not affect the
ratio $\mtherm/\mmin$ as $f \rightarrow 0$.  The same holds true for thermal
fluctuations in the well surrounding \mmin.  

We can rephrase the question of the existence of degenerate
minima as follows: What is the probability that, for a given positive
$E$ and $\epsilon$, 
\be
\cale(m) > \cale(\mmin) + E \;\; \text{for all} \; \;  m \; \; \text{such that}
\; \; |m - \mmin| > \epsilon \Delta/f^2 \; ? \label{inequality}
\ee
If this inequality is satisfied, then $\mtherm/\mmin - 1$ is at most
of the sum of a term of
order $\epsilon$ and of a term of order $\exp(-E/\ktm)$; if for any choice of $E$ and
$\epsilon$ the probability that it is satisfied can be made
arbitrarily close to 1 for $f$ small enough, then Eq.~\parenref{avg-min}
must hold.  One can easily argue from dimensional analysis that this
is the case: The probability \pineq\ that the inequality
Eq.~\parenref{inequality} holds is a function of the dimensionless parameter
$\epsilon$ and of the three parameters $E$, $\Delta$, and $f$, with dimensions,
respectively, of energy, $(\text{energy})^2/\text{nucleotide}$, and
$\text{energy}/\text{nucleotide}$.  Because \pineq\ is
itself dimensionless, it must depend only on dimensionless ratios of
the latter three parameters; by rescaling energies and nucleotide
numbers, one can easily conclude that the only such ratio is $E
f/\Delta$.  Hence, $\pineq = \pineq(\epsilon,E f/\Delta)$.  Moreover, we
know that \mmin\ is the absolute minimum of the random walk, so it must
be true that $\pineq = 1$ when $E = 0$.  As long as $\pineq(\epsilon,E
f/\Delta)$ is a continuous function if its second argument, it must then be true that $\pineq
\rightarrow 1$ as $f \rightarrow 0$ for any fixed $\epsilon$ and $E$.
This is sufficient to confirm that \mmin\ and \mtherm\ coincide with
probability 1 for small $f$.  If $\pineq$ has a well-defined first derivative,
then $1-\pineq \sim E f/\Delta$ for small $f$, a result that can be
verified by a more detailed calculation.  

This linear dependence has a
simple interpretation:  For an unbiased random walk, the
probability to make a first return to the starting point after $m$
steps decays like $1/m^{3/2}$;  this is also approximately the case for a biased random walk on scales
smaller than  $\sim \! \Delta/f^2$.  Upon integrating $1/m^{3/2}$ from
$\epsilon \Delta/f^2$ to some large upper bound, we see that the
probability not to return at all (and thus not to have any minima
nearly degenerate with \mmin) differs from 1 by a number of order $f$.  Our
earlier observation that $\overline{\langle m^2 \rangle - \mtherm^2}
\sim$ goes like $1/f^3$ can also be explained by the small $f$ behavior of
$1-\pineq$~\cite{rand-fp}:  The disorder average is dominated by the probability of
order $f$ that $\langle m^2 \rangle - \mtherm^2$ will be of order
$1/f^4$.  The notion that disorder-averages of higher cumulants can be
determined by rare configurations of the disorder in which there are two
widely-separated minima has been explored in several other random
systems~\cite{dsf-terry,ledou-mon-dsf}.

\subsection{Statistics of Minima: Plateaus and Jumps}
\label{plat-jump}

Having determined that the absolute minimum \mmin\ of $\cale(m)$ and the
average number \mtherm\ of bases opened coincide near the unzipping
transition, we can now use this fact to study the \mtherm\ versus $f$ curve
for a {\em single} random sequence.  Consider the effect on the energy landscape $\cale(m)$
describing a given dsDNA molecule, with a given random sequence,
of tuning the bias $f$ towards zero.  Decreasing $f$ gradually tilts the energy landscape towards the horizontal, as
illustrated in Figure~\ref{energy-tilt}.  The
location of the absolute minimum will then remain constant over a
range of $f$, giving rise to the observed plateaus.  As the landscape
tilts, however, local minima at larger values of $m$ move downwards
faster than those at smaller $m$.  At certain specific values of $f$,
the energy of a minimum at $m > \mmin$ will move below $\cale(\mmin)$,
and $\mmin \approx \mtherm$ will shift from the old minimum
to the new one.  As Figure~\ref{energy-tilt} shows, the two minima can
be separated by a considerable distance, thus giving a physical
explanation for the dramatic jumps seen in Figure~\ref{four-polys}.  

To develop
a quantitative theory of these effects, we begin by calculating
the distribution of \mmin\ for a given $f$, then consider the
conditional probability that $\mmin = m_2$ when $f = f_2$, given that
the minimum was at $m_1$ at a bias $f_1$.  This conditional
distribution will allow us to make predictions, for example, about the
typical sizes of plateaus and of jumps.

We first ask for the
probability $P_{\text{min}}(\mmin,\emin)$ that $\cale(m)$ has its absolute minimum
at $(\mmin,\emin)$, or equivalently the
probability that $\cale(m)$ first reaches \emin\ at ``time'' \mmin,
multiplied by the probability that $\cale(m) > \emin$ for $m > \mmin$.
The latter is simply the splitting probability $\pi$ introduced in the
last subsection.  Although in the continuum approximation $\pi(\cale_0)$ is
singular as $\cale_0 \rightarrow \emin$, we can regularize it in a
manner similar to that used previously.  Because $\pi$ is just a
constant factor, independent of \emin, the details of the
regularization are unimportant.  In practice, $\pi$ can be
determined by demanding that $P_{\text{min}}(\emin,\mmin)$ be
correctly normalized. 

More
interesting is the probability of first passage to \emin.  We first define the probability $S(\cale,m; \emin)$ that, starting from
$\cale = 0$ at $m = 0$, the random walk has arrived at energy $\cale$
after opening $m$ bases, without ever having had $\cale(m) < \emin$.
It turns out that $S$ satisfies the same
Fokker-Planck equation~\parenref{fp-cale} as the probability $P(\cale,m)$
for the unconstrained random walk to arrive at $(m,\cale)$~\cite{vank}.  The
constrained probability $S$, however, is also subject to the boundary
condition $S(\emin,m; \emin) = 0$.  With this boundary condition, one can
solve the Fokker-Planck equation to find
\be
S(\cale,m; \emin) = \frac{1}{\sqrt{2 \pi \Delta m} } \left\{ \exp\left[ -
\frac{(\cale - f m)^2}{2 \Delta m} \right] - \exp\left[ \frac{f
\cale}{\Delta} - \frac{f^2 m}{2 \Delta} - \frac{(2 \emin - \cale)^2}{2
\Delta m} \right] \right\} \; . \label{S-expr}
\ee
The probability to first cross \emin\ after \mmin\ steps is then given
by $(\Delta/2) \partial S/\partial \cale |_{\cale
= \emin}$, i.e. the diffusive flux of
random walkers crossing \emin\ for the first time.  The distribution
 $P_{\text{min}}(\emin,\mmin)$ differs from this function only by a
normalization factor.  Finally, we determine the probability
that the minimum occurs at \mmin\ for any \emin\ by integrating from
$-\infty$ to $0$ with respect to \emin.  The final result is
\be
P_{\text{min}}(\mmin) = \frac{f^2}{\pi \Delta} e^{-m_{\text{min}}
f^2/2 \Delta}
\int_0^\infty d\!w e^{-w m_{\text{min}} f^2/2 \Delta}
\frac{\sqrt{w}}{w+1} \; , \label{min-distr-eqn}
\ee
in agreement with the distribution obtained by le Doussal
{\em et al.} using a real space renormalization
group~\cite{ledou-mon-dsf}.  Note from Fig.~\ref{mindistr-fig} that $P_{\text{min}}(\mmin)$ agrees (to
within counting errors) with the distribution of \mtherm\ obtained from
simulations.  As claimed above, $P(\mmin)$
takes the form of a scaling function of $\mmin f^2/\Delta$.  Variations
in $\mmin \approx \mtherm$ between different random sequences are thus of the same
order as the average \mavg, and the system is not self-averaging.

We now turn our attention to the more interesting and experimentally relevant question of
correlations within a single \mtherm\ versus $f$ curve.  In particular, we
would like to know the probability that $\cale(m)$ has its minimum at
$m_2$ at a bias $f_2$ given that, for the {\em same} realization $\eta(m)$
of the random base sequence, the minimum was at $m_1$ at a bias $f_1 >
f_2$.  This probability will turn out to depend only on the jump size
$\dm \equiv m_2 - m_1$.  The plateaus seen in Fig.~\ref{four-polys} suggest a delta function
contribution at $\dm = 0$.  To determine the strength of this delta
function, consider a polymer with
a fixed base sequence giving rise to an energy landscape 
\be
W(m) \equiv
\int_0^m d\!m' \eta(m') \; . \label{W-defn}
\ee
 If the minimum of $\cale(m)$ is at $m_1$ for
bias $f_1$, then $ W(m) + f_1 m > W(m_1) + f_1 m_1 \equiv \cale_1$ for
all $m$, and hence $ W(m) + f_2 m > W(m_1) + f_2
m_1 \equiv \cale_2$ for $m < m_1$ and $f_2 < f_1$.  If the minimum is
to move from $m_1$ as the bias is tuned down to $f_2$, it must move
towards larger $m$.  This is not surprising---one can easily prove
that $d \mtherm/d f < 0$. 

Let $\Pi_1$ and $\Pi_2$ denote the events,
respectively, that for $m> m_1$, $W(m) + f_1 m > \cale_1$ and $W(m) +
f_2 m > \cale_2$.  The probabilities that $\Pi_1$ and $\Pi_2$ occur
are simply the splitting probabilities $\pi_1 \propto f_1$ and $\pi_2
\propto f_2$.  If the minimum of the random walk falls at $m_1$ for a
bias $f_1$, then $\Pi_2$ is true if and only if the minimum remains at
$m_1$ at the bias $f_2$.  In other words, the coefficient of the delta
function at $\dm = 0$
in the distribution of \dm\ is simply the conditional probability
$\pr[\Pi_2 | \Pi_1]$.  From Bayes' theorem~\cite{feller}, we know that
the probability that events $\Pi_1$ and $\Pi_2$ both occur for the
same random sequence is $\pr[\Pi_2
\wedge \Pi_1] = \pr[\Pi_2 | \Pi_1] \pr[\Pi_1]$.  But if $\Pi_2$
occurs, then $\Pi_1$ must also occur---if the random walk never passes
below its value at $m_1$ with the smaller bias $f_2$, then it can
never do so with the larger bias $f_1$.  Thus, $\pr[\Pi_2 \wedge
\Pi_1] = \pr[\Pi_2]$.  The conditional probability thus takes the simple
form
\be
\pr[\Pi_2 | \Pi_1] = \frac{\pr[\Pi_2]}{\pr[\Pi_1]} =
\frac{\pi_2}{\pi_1} = \frac{f_2}{f_1} \; ,
\ee
and the probability that a plateau stretches from $f_1$ down to $f_2$ is
just $f_2/f_1$.  Upon taking a derivative with respect to $f_2$, we conclude that the end point of a plateau that starts
at a bias $f_{\text{start}}$ is uniformly distributed between $0$ and
$f_{\text{start}}$.  Equivalently, the log ratio $l \equiv
\ln(f_{\text{start}}/f_{\text{stop}})$ of the starting and ending biases
of a plateau is distributed as $\exp(-l)$.

The distribution of plateau lengths, of course, is only part of the
description of a plot of \mtherm\ versus $f$; to complete the characterization, we
must also study the distribution \pj\ of jumps \dm\ for non-zero \dm.
The full distribution of \dm\ will then take the form $(f_2/f_1)
\delta(\dm) + (1-f_2/f_1) \pj(\dm)$.  The calculation of \pj\
requires an extension of our previous first passage approach.  As before, we are interested in the
probability that the biased random walk $W(m) + f_2 m$ first reaches the
energy $\cale_2$ at $m_2 = m_1 + \dm$, but subject now to the additional constraint that
$(m_1,\cale_1)$ is the absolute minimum for the larger bias $f_1$.
Hence, we demand that $W(m) +
f_1 m > \cale_1$ for all $m > m_1$, where $W(m)$ is the same fixed
realization of the random energy landscape.
To calculate this modified first passage probability, note that for each
$m$, only one of the two conditions has to be taken into account.
For $\dm < (\cale_1-\cale_2)/(f_1 - f_2)$, $W(m) + f_1 m > \cale_1$ is
the stronger constraint on the allowed value of $\cale(m)$, while for
$\dm > (\cale_1 - \cale_2)/(f_1 - f_2)$, $W(m) + f_2 m> \cale_2$ is
the stronger.  We can find the first passage probability,
subject to both constraints, by multiplying the probability of
arriving at $\dm = (\cale_1 - \cale_2)/(f_1 - f_2)$ subject to the
first constraint by the probability of going from there to
$(m_2,\cale_2)$ subject to the second.  Specifically, let
$S_1(\cale,m; \emin)$ be the probability of arriving at $\cale$ after
opening $m$ bases, with bias $f_1$ and with $\cale(m)$ always larger
than $\emin$, and let $S_2(\cale,m; \emin)$ be the corresponding
probability with bias $f_2$.  Both probabilities are given by
Eq.~\parenref{S-expr}, with the appropriate substitution for $f$.  As in
our calculation of the distribution of minima, the derivative of $S$
is also an important quantity, so it is useful to define
$S_{1,2}'(\cale,m; \emin) \equiv \partial S_{1,2}/\partial \cale$.
 The
probability that a random walk with bias $f_2$ will arrive at
$(m,\cale)$, subject to the constraint that $W(m) + f_1 m >
\cale_1$, is related to $S_1$ by a ``Galilean'' transformation (with
$m$ viewed as a time and $f_1 - f_2$ viewed as a velocity jump).  Upon making
use of the invariance of the $S\text{'s}$ with respect to uniform
translations in $\cale$ and in $m$, one can thus write the probability
that $(m_2,\cale_2)$ is the minimum at bias $f_2$, given that
$(m_1,\cale_1)$ is the minimum at bias $f_1$, as
\be
P_{\text{jump}}(\cale_2,m_2 | \cale_1,m_1) \propto \int_{\cale_2}^{\infty} d\!\cale'
S_1[\cale' - \cale_1 + (f_1 - f_2) (m'+m_1), m'; 0]
S_{2}'(\cale_2 - \cale',m_2 - m_1 - m'; \cale_2 - \cale') \; , \label{pjump}
\ee
where $m' \equiv (\cale_1 - \cale_2)/(f_1 - f_2)$ is the value of
$\dm \equiv m_2 - m_1$ at which the two constraints switch precedence.  The quantity
$S_1[\cale' - \cale_1 + (f_1 - f_2) m', m'; 0]$, which is formally
zero, is assumed to be regularized by replacing $0$ by $-\epsilon$,
and we have suppressed the normalization factor
proportional to $\pi_2$.  According to Eq.~\parenref{pjump},
$P_{\text{jump}}$ depends only on the two biases $f_1$ and $f_2$ and on the differences \dm\ and $\de \equiv \cale_2 - \cale_1 + (f_1 - f_2) m_1$.  The latter is the
difference between $\cale(m_1)$ and $\cale(m_2)$, both defined with
bias $f_2$; the extra factor proportional to $m_1$ is necessary
because $\cale_1$ is defined with bias $f_1$.  It is straightforward to show that the conditional
distributions of minima are Markovian---that is, the distribution
of $m_2$ and $\cale_2$ does not depend on the location of the absolute
minimum for any $f < f_1$:  Suppose that one were to ask for the
distribution of $m_2$ and $\cale_2$ subject not only to the constraint
that at bias $f_1$, the minimum was at $(m_1,\cale_1)$, but also that
at a bias $f_0 < f_1$, the minimum was at $m_0<m_1$ and $\cale_0$,
with $\cale_1 + (f_0 - f_1) m_0 < \cale_0 < \cale_1 + (f_0 - f_1) m_1$.  This
additional demand translates into the condition that $W(m) + f_0
m > \cale_0$ for $m> m_1$.  This constraint, however, is
weaker than the requirement $W(m) + f_1 m > \cale_1$ imposed by the
location of the minimum at the bias $f_1$.  The distribution of
$(m_2,\cale_2)$ is thus independent of what happens at $f_0$, and the
probability of a given sequence of measurements of $\mtherm = \mmin$
for successive values of $f$ can be expressed as a product of factors
of $P_{\text{jump}}$.

To find the distribution of $\dm$ alone, and thus of $m_2$, one must
integrate $P_{\text{jump}}(\dm,\de)$ with respect to $\de$ from $-(f_1
- f_2) m_2$ to $0$.  The lower bound reflects the constraint that
$W(m_2) + f_1 m_2 > \cale_1$; the upper bound ensures that $\cale_2 <
\cale(m_1)$.  Figure~\ref{jump-distr-fig}
 compares a numerical calculation of the full distribution \pj\
obtained in this way
with simulation results.  The good
agreement confirms that $\mtherm \approx \mmin$.  The
figure also shows that for large \dm, \pj\ decays like
$\exp(-\dm f_2^2/2 \Delta)$.  This is the same as the large \mmin\
behavior of $P_{\text{min}}$ with $f = f_2$; for large enough \dm, the
constraint imposed by the minimum at $m_1$ has no effect on the distribution.

Additional analytic insight can be obtained by considering various limits.  When $(f_1 - f_2)/f_2 \gg 1$, one finds that
$P_{\text{jump}}(\dm,\de) \simeq P_{\text{min}}(\dm,\de)$,
where $P_{\text{min}}$ is the distribution of
the absolute minimum at a given value of $f$ discussed above [Eq.~\parenref{min-distr-eqn}], evaluated with $f = f_2$.  In the
limit of large $f_1 - f_2$, the lower bound on \de\ approaches
$-\infty$, and the integral of \pj\ with respect to \de\ introduces no
extra complications.  The distribution of \dm\ is thus no different
from that of the minimum \mmin\ without any additional constraints.  After
normalization, we find
\be
\pj(\dm) = P_{\text{min}}(\dm) =  \frac{f^2}{\pi \Delta} e^{-\dmm f^2/2 \Delta}
\int_0^\infty d\!w e^{-w \dmm f^2/2 \Delta}
\frac{\sqrt{w}}{w+1} \; \; \; \; \left( \frac{f_1 - f_2}{f_2} \gg 1
\right) \; . \label{jump-large-al}
\ee
Eq.~\parenref{jump-large-al} can be understood as follows:
When $f_2$ is much smaller than $f_1$, the smaller bias allows the
system to visit much more random sequence before the $f m$ term in
$\cale(m)$ makes the energy cost prohibitive.  With so many more
places where the new absolute minimum could occur, the constraint from the
location of the old minimum at the larger bias becomes unimportant,
and $\pj(\dm)$ becomes independent of $m_1$.  Indeed, because $m_2 \sim
1/f_2^2$ is typically much larger than $m_1 \sim 1/f_1^2$, $m_2$
differs very little from $\dm$.  The distribution of $m_2$ thus
approaches $P_{\text{min}}(m_2)$.  That is, the minimum at bias $f_2$
is essentially chosen {\em independently} from the same scaling
distribution as the minimum at bias $f_1$.  With a long enough
sequence, it should in principle be feasible to make many such
independent measurements of $\mmin \approx \mtherm$ at different values of
$f$.  Although DNA unzipping is not self-averaging in the usual
sense, the data from even a single random sequence thus nevertheless
contain remnants of the disorder-averaged behavior.
In particular, if $\ln m$ is plotted versus $\ln f$ for a long enough
polymer, the best-fit line should have slope $-2$, as predicted by our
calculation of the disorder-average \mavg\ (Eq.~\ref{mavg-asympt}), albeit with
considerable scatter about the line.  Figure~\ref{rerand-fig}
illustrates this point.

The distribution of \dm\ in the opposite limit $(f_1 - f_2)/f_2 \ll 1$ is the size distribution of jumps between two successive plateaus, one ending and the other
starting at $f_1 \approx f_2$.  Put in different terms, it gives the
distribution of distances between two essentially degenerate minima at
a given bias $f$, assuming that such minima exist.  Because the two
minima are already required to be at almost the same energy, \pj\ is
independent of \de\ in this limit.  The integral over \de\ is then
elementary, and the resulting distribution takes the form
\be
\pj(\dm) = \frac{f_2}{\sqrt{2 \pi \Delta} } \frac{1}{\sqrt{\dm} }
\exp\left( -\frac{\dm f_2^2}{2 \Delta} \right) \; \; \; \; \left(
\frac{f_1 - f_2}{f_2} \ll 1 \right) \; .
\ee
This expression is valid for $\dm f_2^2/\Delta \lesssim f_2/(f_1 -
f_2)$; for larger values of \dm\, the power law
prefactor of \pj\ crosses over from $1/(\dm)^{1/2}$ to
$1/(\dm)^{3/2}$.  The tail of the distribution thus still agrees
with that of $P_{\text{min}}$, as expected.

Knowledge of \pj\ gives a detailed description of the
statistics of \mtherm\ versus $f$ curves, under the assumption that
\mtherm\ and \mmin\ coincide.  We have already seen that this
assumption is valid with probability 1 as $f \rightarrow 0$.
For any finite $f$, however, there will be occasions when it does
not hold.  In particular, it must break down in the vicinity of jumps between
different plateaus.  Near enough to a jump, the minima giving rise to
the two plateaus will be nearly degenerate, and \mtherm\ will
contain substantial contributions from both minima.  Indeed, if
a jump of size \dm\ occurs at a bias $f_1$, then both minima will be
appreciably occupied if the difference between their energies $|f -
f_1| \dm \lesssim \kt$.  The sharp discontinuity in \mtherm\ at
$f_1$ will be replaced by a smooth transition of width of order
$\kt/\dm$.  We have already seen that \dm\ is typically of order
$\Delta/f_1^2$, so the width of a typical transition sharpens like
$f_1^2$.  In contrast, we have seen that a typical plateau at bias $f_1$
extends for a distance of order $f_1$.  As $f_1 \rightarrow 0$, the width of the
jumps thus becomes very small compared to the size of the plateaus, in
agreement with our arguments that $\mmin/\mtherm \rightarrow 1$ in this
limit.  The sharpening of the jumps as $f\rightarrow 0^{+}$ is evident
in Figs.~\ref{four-polys} and ~\ref{rerand-fig}.

Note also that if the temperature is raised
at fixed force near the unzipping transition (i.e. a vertical instead
of a horizontal
trajectory in the inset to Fig.~\ref{fig1}), we have $f \sim T_{\text{C}} -
T$.  The surface contribution to the specific heat near the transition
is thus $T
\partial^2 G/\partial T^2 \sim T^2 \partial \ln Z/\partial f^2 \sim
\partial \mtherm/\partial f$, where $G = -\kt \ln Z$ is the ``surface''
free energy of the partially unraveled polymer duplex at fixed temperature and force defined in
section~\ref{non-rand-sect}.  If \mtherm\ as a function of $f$ takes
the form of a sequence of plateaus and jumps, then the derivative of
\mtherm\ with respect to $f$ must vanish except in the vicinity of the
jumps, where it will show a sharp spike proportional to the jump size
\dm.  As $f \rightarrow 0$ and the jumps become very sharp, the specific
heat spikes will approach delta functions.  Each jump can thus be
thought of as a ``micro-first-order transition.''

We close this subsection with an example of how plateaus and jumps can
appear in the unzipping of a biologically relevant DNA sequence, that
of phage lambda~\cite{gen-bank}.  Figure~\ref{lam-en} plots the energy
landscape $\cale(m)$ of a 28
kb segment of the lambda genome for two different
biases.  The energy to
open each base pair is taken from a widely-used parameter
set~\cite{dna-en}, and we neglect
the possibility of rare denaturation bubbles under physiological
conditions.  The energy landscape shows two pronounced minima; a third
minimum very near $m = 0$ is barely visible.  The corresponding plot
of \mtherm\ versus the distance $f \sim \fc - F$ from the
transition, determined by an exact evaluation of the partition
function, appears in Figure~\ref{lam-opening}.  As expected, it
consists of three plateaus, corresponding to the three minima.  Thus, the qualitative ideas developed in this section apply to
real sequences found in experimental biology as well as to the
idealized random models explored here.

\subsection{Application: Determination of base-pairing energies.}

In this subsection, we digress briefly from our primary focus on
polynucleotides with random sequences to discuss how the
mechanical denaturation of specially designed sequences might be used to measure the strength of the
base-pairing and stacking interactions that stabilize polynucleotide
duplexes.  Traditionally, these interactions have been studied by
analyzing the thermal melting curves of double-stranded DNA's and
RNA's~\cite{dna-en,polynucl-en}.  Most commonly, the
stability of a duplex is assumed to be determined by 10 phenomenological
parameters giving the combined pairing and stacking energies of the 10
possible distinct groups of two successive base pairs.  These
parameters can be inferred from the melting temperatures of a set of
duplexes with appropriately chosen sequences.  Although in most ways
quite successful, this method has the disadvantage that it yields
values the 10
energy parameters only in the vicinity of the melting temperatures of
the double-stranded molecules.  Because these energy parameters are
expected to depend on a variety of conditions, including salt
concentration, pH, and (for entropic reasons) temperature, it would be useful to have a
technique that allowed the measurement of duplex stability in a wider
range of conditions.  It has already been shown experimentally that
micromechanical experiments can be used to estimate the binding energy
of a particular RNA hairpin~\cite{liphardt}.  Here we extend the
analysis of~\cite{liphardt} to consider more generally how mechanical
denaturation might be used to infer the stability of duplexes.

Because it is difficult to synthesize long polynucleotides with
prescribed sequences, one would like to be able to measure pairing energies
on relatively short (tens of nucleotide) hairpins.  Even for short
hairpins, one can still define an average pairing energy $g_0$, a
variation about the average $\eta(m)$, a critical unzipping force \fc\
satisfying Eq.~\parenref{fc-defn}, and a distance $f = 2 g(\fc) - g_0$
from the transition.  Drawing on the ideas developed in
subsections~\ref{zero-temp-fp} and~\ref{plat-jump}, we expect that, for a
given hairpin in the constant force ensemble, \mtherm\ will remain
close to minima of $\cale(m)$ except for jumps at certain values of
$f$.  The measurement of $g_0$ for a hairpin of length $N$ is most
straightforward if there are only two such minima, at $m = 0$ and $m =
N$; the unzipping transition then shows two-state
behavior~\cite{dna-en}.  For this to be the case, the energy landscape
$\cale(m)$ must take roughly the form shown in Fig.~\ref{enmeas-tilt}.  Because
of the energy barrier between $m = 0$ and $m = N$, the unzipping fork
is always localized in the vicinity of one of these minima, with a
sharp jump between the two at \fc\ (Fig.~\ref{enmeas-open}).  \fc\ is thus easily
read off from the experimental extension versus force curve, and $g_0$
is then given by Eqs.~\parenref{g(F)-gen} and~\parenref{fc-defn}.  Just as for the standard
methods based on melting curves, the 10 energy parameters can be
estimated from the knowledge of $g_0$ for enough different hairpins.
One can straightforwardly design hairpins with two-state unzipping
behavior by joining a stretch of strongly-paired bases to a less
stable stretch.  Thus, for example, if one strand of the hairpin has
sequence $5'(\text{C})_{N/2}(\text{A})_{N/2}3'$, with opening starting
from 5' end (and complementary sequence $3'(\text{G})_{N/2} (\text{T})_{N/2}5'$), $g_0$ for the hairpin approaches for large $N$
the average of the energies associated with (reading along one strand
of a duplex) $5'\text{CC}3'$ and $5'\text{AA}3'$.  Similarly,
$5'(\text{CG})_{N/4}(\text{AT})_{N/4}3'$, paired with its complement, gives the average of the energies
associated with $5'\text{CG}3'$, $5'\text{GC}3'$, $5'\text{AT}3'$, and
$5'\text{TA}3'$.  Corrections due to the junction between
the two homopolymeric stretches and to the confinement energy of the
loop section of the hairpin both decay like $1/N$; they can be
eliminated by measuring hairpins with several different values $N$.  Mechanical
denaturation in the constant force ensemble can thus be used
systematically to determine the 10 standard duplex stability
parameters in a wide range pH, salt concentration, and temperature.

\section{Constant Extension Ensemble}
\label{const-ext-sect}

So far, we have considered only the constant force ensemble, in which
a fixed force is applied to the two single strands of the dsDNA,
and one measures the average number of base pairs opened or the average separation
$\langle \rb \rangle$ between the ends of the two single
strands. Constant extension experiments, in which the separation \rb\ is fixed, and the average force
is measured, are also possible.  In the classical thermodynamics of macroscopic systems,
these two ensembles would be equivalent.  That is, the functions
$\langle \rb \rangle(\fb)$ and $\langle \fb \rangle(\rb)$ measured in the two
ensembles would be inverses of each other.  In single
molecule experiments, however, such a relation is not guaranteed, and
the two ensembles are in fact not
equivalent in DNA unzipping.
For simplicity, we assume throughout this section that the Kuhn
length $b$ of the single-stranded polymer is equal to the length $a$
per chemical monomer.

We begin by considering the constant extension ensemble in the {\em absence}
of sequence randomness.  We neglect long-ranged
interactions within the single-stranded polymers; because \rb\ and
\fb\ will always be parallel on average, we can work with the (signed)
scalars $r$ and $F$.  Regardless of the elastic properties of the
single stranded DNA (freely-jointed chain, Gaussian, etc.), one can define the statistical weight $G_{2
m}(r)$ for a single-stranded chain of length $2 m$ to have an
end-to-end distance $r$.
The partition function \zext\ in the constant extension ensemble can
then be viewed as a weighted sum over the number of unzipped bases $m$
with $r$ fixed. Given the energy cost $g_0 m$ of opening $m$ bases, one
has~\cite{terry-uli,heslot} 
\be
 \zext(r)  =  \int_0^{\infty}\! d\!m \, G_{2 m}(r) \exp(-g_0 m/\kt) \; .
\ee
In the limit of large $r$, one expects the number of unzipped bases
$m$ to be proportional to $r$.  It then makes sense to consider the
free energy per base $h(x)$ of the liberated single strands as a
function of the extension per base $x \equiv r/2 m$. The free energy per base $g(F)$ in the constant force
ensemble is related to $h(x)$ by the Legendre transform $g(F) =
h[x(F)] - F x$, and in the thermodynamic limit $r
\rightarrow \infty$ with $r/m$ fixed, we expect $-\kt \ln[G_{2 m}(r)] \simeq 2 m
h(x)$.  It is not difficult to show that the leading correction to
this result is of order $\ln(m)/2$.  Hence, for large $r$ the partition
function becomes, up to $r$-independent multiplicative constants,
\bea
\zext(r) & \simeq & \int_0^{\infty} \frac{d\!m}{m^{1/2}} \exp\left[-\frac{2
m}{\ktm} h\left(\frac{r}{2 m}\right) - \frac{g_0 m}{\ktm} \right]
\nonumber \\
& = & \sqrt{\frac{r}{2} } \int_0^{\infty} \frac{d\!n}{n^{1/2}}
\exp\left\{ - \frac{r}{\ktm} \left[ n h(1/n) -
\frac{g_0 n}{2} \right] \right\} \; , \label{zext-with-x}
\eea
where we have introduced $n =
2 m/r = 1/x$ in the second line.  For large $r$, \zext\ may be evaluated in the saddle point
approximation, which gives
\be
\zext(r) \simeq \sqrt{ \frac{ (n^*)^2 \pi}{h''(1/n^*)} } \exp\left[
-\frac{r h'(1/n^*)}{\kt} \right] \left[ 1 + {\cal
O}\left(\frac{1}{r}\right) \right]\; . \label{zext-with-saddle}
\ee
The ${\cal O}(1/r)$ term comes from subleading corrections to
$G_{2 m}(r)$ that we have chosen not to calculate explicitly.
The location $n^*$ of the saddle point satisfies $h'(1/n^*) = (n^*) [h(n^*)
+ g_0/2]$, where $h'(x) \equiv dh/dx$ plays the role of a force.
Indeed, upon using the Legendre transform
relation between $h$ and $g$, we find that $h'(1/n^*)
= \fc$.  Thus, for large $r$, the average force in the constant extension ensemble
takes the simple form
\be
\langle F \rangle = -\kt \frac{\partial \ln \zext}{\partial r} \simeq
\fc + {\cal O}\left(\frac{1}{r^2}\right) \;.
\ee
In the constant force ensemble, on the other hand,
$\langle r \rangle \propto 1/f \sim 1/(\fc - F)$, which upon inversion
gives the slower approach to \fc\ $F = \fc + {\cal O}(1/r)$. Both
ensembles predict that complete unzipping of the dsDNA occurs at
$F = \fc$; in fact, in the limit $r \rightarrow \infty$, the constant extension
ensemble simply demonstrates coexistence of the bulk unzipped and
zipped phases, as in any first order transition.  The
approach to $F = \fc$ as $r$ becomes large, however, is markedly different.  Equivalence of ensembles exists only in the
``thermodynamic limit'' $r \rightarrow \infty$.

Because DNA unzipping does not show self-averaging, the situation
becomes even more complicated when sequence randomness is introduced.
In the constant force ensemble, \mtherm\ (and hence $\langle r
\rangle$) increases monotonically as $F$ increases, for any DNA
sequence.  In the constant extension ensemble, in contrast, we expect large regions where $d \cale/dm$, which
plays roughly the role of $g_0$, is smaller than average; when the
unzipping fork enters one of these regions, $\langle F \rangle$ should
{\em decrease}.  Precisely such behavior is observed in experiments
and simulations on the unzipping of lambda phage DNA~\cite{heslot}:
$\langle F \rangle$ is seen to vary randomly about an average value as
$r$ is increased.  For a given random sequence, the functions $\langle
r \rangle(F)$ and $\langle F \rangle(r)$ thus cannot be inverses of
each other.  

One can still ask, however, whether the disorder averages
$\ravg(F)$ and $\favg(r)$ are simply related.  Once sequence
heterogeneity is present, a term proportional to $W(m)$ [see
Eq.~\parenref{W-defn}] must be incorporated into $\zext(x)$.  In analogy to
Eqs.~\parenref{zext-with-x} and~\parenref{zext-with-saddle}, one finds
\bea 
\zext(r) & \simeq & \int_0^{\infty} \frac{d\!m}{m^{1/2}}
\exp\left[-\frac{2 m}{\ktm} h\left(\frac{r}{2 m}\right) - \frac{g_0
m}{\ktm} - \frac{W(m)}{\ktm} \right] \\ \nonumber
 & \simeq & \sqrt{r} \exp\left[
-\frac{ x \fc}{\ktm} - \frac{\sqrt{r} W(n^*)}{\sqrt{2} \ktm} \right]
\int_{-\infty}^{\infty} \frac{d\!n}{(n)^{1/2} } \exp\left[-\frac{r k
(n-n^*)^2}{2 \ktm} - \frac{\sqrt{r} W(n - n^*)}{\sqrt{2} \ktm} \right]
\;, \label{randz-saddle} 
\eea 
where $k = (1/n^*)^3 h''(1/n^*)$.  In
passing from the first to the second expression, we have used the
scaling properties of a random walk to make the substitution, valid on
the level of statistical distributions, $W(r n/2) = \sqrt{r/2} W(n)$.
We have also expanded around the location $n^*$ of the saddle point in
the {\em non-random} case.  Because the average terms in the exponential
grow like $r$, while the coefficient of $W(n)$ is only proportional to
$\sqrt{r}$, this expansion will still give the correct asymptotic
behavior as $r \rightarrow \infty$.

Eq.~\parenref{randz-saddle} shows that the leading corrections to
$\favg(r)$ can be described by the equilibrium extension of a spring
``dragged'' across a random potential~\cite{terry-uli}.  One can
estimate the spring's extension by balancing the elastic energy cost
of extension $-r k (n - n^*)^2/2$ with the typical random energy gain $
\sqrt{r} W(n - n^*) \sim \sqrt{r \Delta |n-n^*|}$.  These two terms are
of the same order when $ |n - n^*| \sim (\Delta/k^2 r)^{1/3}$.  The
typical energy gain due to extension is then $ \sqrt{r \Delta (n -
n^*)} \sim (\Delta^2 r/k)^{1/3}$; note that although $n-n^*$ is positive
or negative with equal probability, the associated change in energy
must always be negative. We thus expect that the disorder-averaged
free energy should behave like
\be
-\kt \overline{\ln \zext(r)} \sim r \fc - \left( \frac{\Delta^2 r}{k}
\right)^{1/3} \; .
\ee
Note that the term proportional to $W(n^*)$ averages to zero.  Upon taking a
derivative with respect to $r$, one concludes that the disorder-averaged
force in the constant extension ensemble approaches \fc\ for large $r$ according to
\be
\fc - \favg(r) \sim \left( \frac{\Delta^2}{k} \right)^{1/3}
\frac{1}{r^{2/3} } \; .
\ee
In contrast, in the constant force ensemble, $\ravg \sim \mavg \sim
1/(\fc - F)^2$, which upon inversion gives $\fc - F \sim
1/\ravg^{1/2}$.  Once again, the two ensembles agree {\em only} on the
location of the unzipping transition!

There is one further, more subtle relationship between the two
ensembles with sequence randomness.  For a given sequence, the
constant force partition function can be written in either of two
ways:
\be
Z(F) = \int_0^{\infty} d\!m \exp\left[ -\frac{W(m) + f m}{\ktm}
\right] = \int_{-\infty}^{\infty} d\!r \exp\left[ -\frac{ \df(r)+ (\fc
- F) r }{\ktm} \right] \;,
\ee
where $\df(r) \equiv -\kt \ln[\zext(r)] - \fc r$.  These two
expressions must of course ultimately lead to the same result, and
this fact has interesting consequences for the properties of \df.
Near the unzipping transition, $f = 2 g(F) - g_0 \sim 2 g'(\fc) (F -
\fc) = 2 |g'(\fc)| (\fc - F)$.  Up to a constant
factor (and neglecting exponentially suppressed contributions to the
second integral from $r < 0$), both expressions for $Z(F)$ can thus be
viewed as Laplace transforms with respect to the same variable.
Hence, we expect that $\df(r)$ must have statistics very
similar to those of $W(m)$.  In particular, for small $\fc - F$, the
integral with respect to $r$, like the one with respect to $m$, is
likely to be dominated by its absolute minimum.  In order to give the
correct sequence of plateaus and jumps, $\df(r)$ should thus behave
like a random walk for large $r$, with $\overline{(\df(r') -
\df(r))^2} \simeq 2 |g'(\fc)| \Delta |r' - r|$.  Scaling arguments due
to Gerland {\em et al.}~\cite{terry-uli} suggest that the force
deviation $\delforce(r) \equiv \langle F(r) \rangle -
\fc = \partial \df/\partial r$ should have a variance that decays like
$\overline{ \delforce(r)^2} \sim \Delta^{1/3} (k/r)^{2/3} $.  For
$\overline{(\df(r') - \df(r))^2}$ to behave correctly at large scales,
$\delforce(r)$ must then have a correlation length that grows like
$r^{2/3}$.  One plausible explanation for this behavior is that, much
as in the constant force ensemble, $n$ locks into a single minimum of
$W(n)$ as $r$ is increased over a finite interval before jumping to a
new minimum, with the size of this interval increasing as $r$ grows larger.

\section{Dynamics}
\label{dynamics}

So far, we have only considered static, equilibrium behavior.  In real
experimental systems, of course, dynamical effects can play an
important role.  The complete description of the dynamics of the
unzipping transition, allowing for the possibility of thermally
activated denaturation bubbles in the bulk dsDNA, is a challenging and
still open problem.  Four time scales come into play: the time scales
\tend\ and \tbulk\ of base pairing and unpairing at the end of a
double-stranded region and in the bulk, the relaxation time $\tss(m)$
of the liberated single strands, and the rotational relaxation time
$\trot(m)$ of the still zipped dsDNA, which because of its helical
structure develops excess twist as it is unravelled from one end.  The
latter two time scales are expected to depend on $m$.
Cocco {\em et al.}~\cite{cocco-marko} have suggested that there may be
a fifth scale associated with overcoming an additional energy barrier
to unzipping the first few bases of an initially blunt-ended dsDNA,
but such a barrier would not affect the long time unzipping dynamics.
Although not the subject of extensive investigation, the opening rate
\tend\ of terminal base pairs is thought to be between 1 and 10
msec~\cite{cocco-marko,libch-group}.  Because opening a base pair in the middle of a
double-stranded region requires overcoming {\em two} stacking
interactions, instead of one for opening at the end, we expect $\tbulk \gg \tend$~\cite{cocco-marko,frank-k};
in unzipping experiments, the pulling force will further accelerate
base-pair opening at the unzipping fork.  Marenduzzo {\em et
al.}~\cite{maritan2} have argued that the relaxation time of the ssDNA
is given by the time required to move the entire single strand a
distance $x$ for each monomer that is opened or closed.  Because the
forces required to denature dsDNA are fairly large, each
single-stranded monomer will be under considerable tension, with the
average extension $x$ per monomer of order the monomer size $a$.  The
mobility of a single strand of length $m$ is then of order $1/(4 \pi
\eta a m)$, where $\eta$ is the solvent viscosity, regardless of
whether the strand is described by the Rouse or by the Zimm model.
Assuming a force $F$ of order 10 pN, one then finds that $\tss(m) \sim
4 \pi \eta a^2 m/F \sim (1 \text{nsec}) m$.  Similarly, we can
estimate the rotational relaxation time \trot\ of a dsDNA molecule of
length $N-m$ by finding the time for it to turn through $2\pi/10.5$ radians (with the denominator of 10.5 arising from the number
of base pairs per helix turn in B form DNA in solution~\cite{klug}).
For a dsDNA strand of radius $1 \text{nm}$, the torque
exerted by the two single strands under tension is roughly $2 \times
10 \text{pN} \times 1 \text{nm} = 20 \text{pN}\cdot \text{nm}$.
Classically, the rotational mobility \rotmob\ of a dsDNA molecule of
length $N$ has been calculated by assuming it is a straight, rigid
rod, yielding the value $\rotmob \approx (2 \times 10^{-8}
\text{sec}/\text{g}\text{cm}^2) N$~\cite{levinthal}; this would imply
$\trot \sim (3 \text{nsec}) (N-m)$.  More recently,
Nelson has argued that the presence of intrinsic bends in natural
dsDNA could decrease the rotational mobility, and thus increase \trot,
by several orders of magnitude~\cite{pnelson}.

The time dependence of the number of unzipped bases $m(t)$ will be
determined by which of these four time scales is the slowest.  The
most difficult situation to analyze occurs if the system is dominated
by \tbulk, as we expect to be the case for small enough $m$ and $N$.  In this case, the dynamics of the
denaturation bubbles in the bulk dsDNA will be slower than the
dynamics of the actual unzipping.  Unlike in our equilibrium
calculations, the bubbles then cannot be integrated out to give an
effective (local) dynamics that depends only on $m$.  Indeed, in the
limit that bases at the unzipping fork open much faster than those in
the bulk, the unzipping fork will propagate into an almost frozen
landscape of opened and closed base pairs.  Strongly non-equilibrium effects, including a
depression of the effective \fc, could then become
apparent~\cite{dsf-personal}.  Because $\tss(m)$ grows with $m$, it
must eventually become slower than \tbulk; beyond this point, more
conventional behavior should reemerge.

Fortunately, in physiological conditions, opening of base pairs in
bulk dsDNA is extremely rare.  Well below the melting temperature, it
is then reasonable to assume that all base pairs beyond the unzipping
fork are closed, and to focus only on the position of the unzipping
fork.  Consider first the case in which the slowest of the three remaining
timescales is independent of $m$, either because \tend\ is the slowest
(as will be the case for $m < N \lesssim 10^3$ or $10^4$ under the
assumption of a straight rod rotational mobility for dsDNA) or because
\trot\ is the slowest, but with $m \ll N$ so that changes in $m$ have
a negligible effect on \trot.
In this regime, the unzipping
dynamics is dominated by the diffusion of the unzipping fork in the
one-dimensional energy landscape $\cale(m)$.  In other words, it is an
example of the well-studied problem of a random walk in a random force
field, sometimes known as the Sinai problem~\cite{sinai} (for
reviews, see~\cite{bouch-geo}).  The overdamped dynamics
associated with the continuum free energy of Eq.~\parenref{cale-defn}
then takes the form
\bea
\frac{d m}{d t} & = & - \Gamma \frac{\delta \cale(m)}{\delta m} +
\zeta(t) \nonumber \\
& = & - \Gamma [f + \eta(m)] + \zeta(t) \; , \label{cont-sinai}
\eea
where the effect of thermal fluctuations is included through the noise
source $\zeta(t)$ with correlations
\be
\langle \zeta(t) \zeta(t') \rangle = 2 \kt \Gamma \delta(t - t') \;
. \label{zeta-defn}
\ee
The magnitude of the phenomenological drag coefficient $\Gamma$ is set
by the slowest time scale:
\be
\Gamma = \frac{1}{\tau \ktm} \; ,
\ee
with $\tau$ equal to \tend\ or \trot\ as appropriate.
We expect that Eq.~\parenref{cont-sinai} describes the dynamics of the
unzipping fork at long times for small $f$.  In the absence of
sequence hetergeneity ($\eta(m) = 0$), it yields simple diffusion with
drift above the unzipping transition, 
\be 
\langle m(t) \rangle = (\Gamma |f|) t \; \; \; \; \text{and} \;\;\;\;
\langle[ m(t) - \langle m(t) \rangle ]^2 \rangle = (\Gamma \kt) t \; .
\ee
In contrast, in the presence of sequence heterogeneity, the
long time dynamics is determined by large energy
barriers that grow with $m$; a number of rigorously-established
results can then be reproduced by simple physical arguments~\cite{bouch-geo,ledou-mon-dsf,dsf-personal}.  For example, when $F = \fc$ (i.e. $f = 0$), $\cale(m)
\sim \sqrt{\Delta m}$; taking this to be a typical barrier size, one finds that the time to go a distance $m$ is $t
\sim \tau \exp(\sqrt{\Delta m}/\kt)$, suggesting that $m(t)$ is
typically of order $\ln^2(t/\tau)$.  Indeed, it is known that in the
presence of a single reflecting wall (in our case, the end from which the
semi-infinite duplex is being unzipped), the ratio $m(t)/\ln^2(t/\tau)$ approaches a
$t$-independent limiting distribution at large times~\cite{ledou-mon-dsf}.
Similarly,
just {\em below} the unzipping transition, the unzipping fork is essentially
always in a region where the small bias $f$ can be ignored.  Given
that $\overline{\dm} \sim \mavg \sim \Delta/f^2$, we expect that the typical time to equilibrate at a bias $f$ (and in particular to jump
from one local minimum to a new minimum with lower energy as $f$ is
decreased) should be of order $\tau \exp(\sqrt{\Delta}/f)$, a
result that is supported, up to logarithmic factors, by
renormalization group calculations~\cite{ledou-mon-dsf}. 

 Just {\em above}
\fc, the dsDNA must eventually unzip completely, but the propagation
of the unzipping fork is again dramatically slowed by the presence of large energy
barriers.  The distribution of barrier heights is known to have exponential
tails~\cite{derrida}, leading to a distribution of trapping times $T$ that decays
like $1/T^{\mu + 1}$, with 
\be
\mu = 2 \kt |f|/\Delta \; .
\ee  
This same exponent
appeared, for example, in Eq.~\parenref{soln-x}, and is known more generally to
control the probability of large excursions of a biased random walk
[e.g. $\cale(m)$] against its bias~\cite{altschul}.  The time to open
$m$ base pairs is a sum of ${\cal O}(m)$ such trapping times, with
each time chosen independently.  For $\mu < 1$, the median value of
this sum grows like $m^{1/\mu}$, so one has sublinear growth with time
of the
sequence-averaged degree of unzipping,
\be
\overline{\langle m(t) \rangle} \sim t^{\mu} \;\; \; (\mu < 1) \; .
\ee
The average extent of unzipping $\langle m(t) \rangle$ of a given
polynucleotide is typically also of order $t^{\mu}$, but with time and
sequence-dependent fluctuations in the prefactor.  For $1 < \mu < 2$,
$ \overline{\langle m(t) \rangle} \sim t $ recovers its usual
behavior, but there is still anomalous behavior in the second
cumulant: $\langle m(t)^2 \rangle - \langle m(t) \rangle^2$ typically
grows like $t^{2/\mu}$.  Conventional diffusion with drift is
recovered only for forces large enough that $\mu > 2$, or $|f| >
\Delta/\kt \sim {\cal O}(\kt)$ for dsDNA in physiological conditions.
For the freely-jointed chain expression~\parenref{fjc} for $g(F)$,
this condition translates to $F - \fc \gtrsim 5 \text{pN}$; there is
thus a substantial window where anomalous drift can be observed in a
single molecule experiment.  Just as for the equilibrium results
discussed earlier in this paper, most of the qualitative features of
the unzipping dynamics for uncorrelated random sequences also apply to
the unzipping of correlated random sequences, albeit with different
exponents~\cite{fei-vin}.

These results have interesting implications for attempts to read
sequence information via experiments which monitor the velocity
$dm/dt$ of the unzipping fork for a fixed force $F > \fc$.  Read
naively, Eq.~\parenref{cont-sinai} suggests that the coarse-grained
sequence fluctuations embodied in $\eta(m)$ and the thermal noise
$\zeta(t)$ will together modulate a mean unzipping velocity
$\overline{\langle dm/dt \rangle} = \Gamma |f|$.  This picture is
certainly correct sufficiently far above the unzipping transition,
where deep traps in the energy landscape are rare.  However, we can
estimate that thermal fluctuations will the obscure the
sequence-dependent modulation of the mean velocity whenever
\be
\langle \zeta(t) \zeta(t') \rangle \gg \Gamma^2
\overline{\eta(\Gamma |f| t) \eta(\Gamma |f| t')}  \; ,
\ee
where we have used the zeroth-order relation 
\be
m(t) \approx \Gamma f t \label{drn2}
\ee
to approximate $m(t)$.  Eqs.~\parenref{eta-defn} and~\parenref{zeta-defn}
then show that thermal noise can only be neglected provided
\be
2 \kt \Gamma \ll \frac{\Gamma \Delta}{|f|} \; ,
\ee
or for
\be
\mu = \frac{2 \ktm |f|}{\Delta} \ll 1 \; .
\ee
In this regime, however, the approximation~\parenref{drn2} breaks
down; indeed, we have seen that for $\mu < 1$, the dynamics is dominated by the
presence of deep traps in the energy landscape, with $m(t) \sim t^{\mu}$.  Efforts to
extract sequence information from $dm/dt$ in this regime will be
seriously hampered by the slow, erratic dynamics associated with
energy barriers of order $\sqrt{\Delta m}$.

The results discussed above are valid as long as the slowest time
scale $\tau$ is roughly independent of $m$.  If $m$-dependence become
important, large energy barriers still dominate the dynamics, but our
arguments must be modified to account for this new
feature~\cite{maritan2}.  Thus, for example, if $m$ becomes large
enough, the relaxation of the single strands will set the basic scale
for the dynamics.  Exactly at the transition, we then expect $t \sim m
\exp(\sqrt{\Delta m}/\kt)$ (the prefactor of $m$ arising from the fact
that $\tss \sim m$); this yields exactly the same very slow asymptotic
behavior $\mavg \sim \ln^2(t)$ as before.  Likewise, the equilibration
times below the transition remain unchanged.  On the other hand, for
$F > \fc$, new behavior emerges.  The time to go a distance $m$ is now
of order $ \sum_{n = 0}^m n T_n$, with each of the $T_n$ chosen from
the same distribution with tails like $1/T_n^{\mu + 1}$.  The median
of the distribution of this new sum occurs at a time of order $m^{(\mu
+ 1)/\mu}$, suggesting $\mavg(t) \sim t^{\mu/(\mu + 1)}$.  As
hypothesized in~\cite{maritan2}, the scaling laws in this regime are
thus related to those for $\tss(m) < \tend$ by the substitution $t
\mapsto t/x$.  Similarly, when $\trot(m) \sim N - m$ is the slowest
timescale, the logarithmic growth at or below the transition remains
unchanged, while above the transition an analysis of a sum of trapping
times $\sum_{n=0}^m (N-n) T_n$ suggests $\mavg(t) \sim N [1 - (1 - k
t^{\mu}/N^{\mu + 1})^{1/(\mu+1)} ]$, with $k$ an undetermined
constant.  Thus, the fact that $\tss$ and $\trot$ depend on $m$ does
not change the essential physical result that sequence randomness
leads to large energy barriers, and thus to a substantial slowing down
of unzipping.

\section{Conclusions}
\label{conclusions}

In this paper, we have given a detailed theoretical analysis of a
simple micromechanical experiment: the mechanical denaturation, or unzipping, of
double-stranded DNA with a random base sequence.  Although of current
experimental interest in its own right, this system can also
serve as a springboard for developing ideas with potential applications to
micromanipulation experiments on more structurally complicated
biomolecules.  Several such ideas emerged from our study.  On the most
basic level, the constant force and
constant extension ensembles were shown to give different force-extension curves in single molecule experiments.
We argued that unzipping in the constant force ensemble can always be
described by a one-dimensional free energy landscape $\cale(m)$, with
an average slope $f = 2 g(F) - g_0$ set by the applied force $F$ and
$F$-independent fluctuations about this average determined by the
structure and sequence of the molecule being examined.  The number of
monomers $\langle m \rangle$ liberated at a given $F$ is then simply
an equilibrium average over $m$ with weight $\exp[-\cale(m)/\kt]$.  Once
sequence variation is present, $\cale(m)$ will in general pass below
zero for small enough $f>0$.  Partial mechanical denaturation then
allows the liberated monomers to gain more free energy by
aligning with the applied force than they lose by breaking native
contacts.  For small $f$, \mtherm\ should be
dominated by the deepest minima in $\cale(m)$.

For the particular case of unzipping a single dsDNA molecule, these
qualitative observations can be given more precise meaning. The energy function
$\cale(m)$ then behaves like a biased random walk on scales beyond
a few bases.  When the
pulling force $F$ is increased to a critical value \fc, the bias $f$
changes sign, and a phase transition occurs.  Randomness is always
relevant at this wetting-like transition, with the average number of
broken base pairs \mtherm\ diverging like $1/(\fc - F)$ for
homopolymer duplexes, but like $1/(\fc - F)^2$ in the presence of a random
sequence.  Individual dsDNA molecules approaching the unzipping
transition open in a sequence of sharp jumps, separated by long
plateaus in which $\langle m \rangle$ remains essentially constant.
The jumps become sharper and sharper as $f \rightarrow 0$.
For small $f$, \mtherm\ for any given polymer must
approach the absolute minimum of $\cale(m)$.  The plateaus and jumps
can then be understood as arising from a sequence of minima.  A given
minimum remains stable over a range of $f$ values.  As the bias $f$
decreases, however, eventually a new minimum at larger $m$ will become
lower in energy; at this point, \mtherm\ will jump to the new minimum.
Starting from this picture, we were able to make precise predictions about
statistical features of single molecule unzipping such as the
distribution of jump sizes \dm.  These showed good agreement with
simulations.  The distribution of \dm\ also revealed that the
correlation between \mtherm\ at different values $f_1$ and $f_2 < f_1$ of
$f$ vanishes for small $f_2/f_1$.  As a result, even though \mtherm\
can differ significantly from \mavg\ at any single force value, a plot
of \mtherm\ versus $f$ for a given random sequence still shows the
same scaling behavior as does the average over many sequences \mavg.
Several of these features, most notably the dominance of the absolute
minimum, are known to occur more generally in random systems; indeed,
an added interest of DNA unzipping is that it is a physical
realization of one of the simplest models in the statistical mechanics
and dynamics of random
systems~\cite{sinai,bouch-geo,rand-fp}.  Similar conclusions should apply to experiments
on the
unzipping of individual RNA hairpins~\cite{liphardt}, although
experiments on longer hairpins would be required to provide a complete
test of the theory.

Although the predictions for DNA unzipping do not apply directly to
micromechanical assays on systems such as proteins~\cite{protein-pull}
or the complex RNA folds of naturally-occuring ribozymes~\cite{liphardt}, they do suggest a definite agenda for
understanding such experiments.  In varying the pulling force $F$ in
the constant force ensemble, one is essentially searching for local
minima along the denaturation pathway; each observed plateau
corresponds to a state that is metastable at zero force, but is
stabilized in an appropriate range of $F$ values.  If $g(F)$ can be
determined from measurements on unfolded strands, then the energies of
the original metastable states are easily inferred from the forces at
which jumps occur.  Related ideas have been applied with great success
to
the interpretation of micromanipulation experiments on individual ``lock
and key'' bonds~\cite{evans}.

This picture of plateaus and jumps can could break down if, instead of
traversing only a single pathway,
the mechanical denaturation can proceed along one of many different
routes~\cite{terry-uli}.  For example, in micromanipulation experiments
on folded RNA's, it can transpire that a series of many hairpins are
under tension simultaneously, as in Figure~\ref{rna-fig}.  In the
constant force ensemble, if there are $M$ long hairpins with independently
chosen random sequences, the average extension \ravg\ will be simply
the sum of $M$ independent single hairpin extensions.  As a function
of $f$, each of these single hairpins will go through its own sequence
of plateaus and jumps.  Each time a particular hairpin has a jump,
\rtherm\ will also jump, but the typical jump size will be $\rtherm/M$
instead of \rtherm.  Similarly, the plateaus in \ravg\ will be
shortened: The probability that none of the single hairpins jump as
$f$ is decreased from $f_1$ to $f_2$ is $(f_2/f_1)^M$, which decays
very quickly for large $M$.  As $M$ increases, shorter and shorter
jumps and plateaus will eventually merge into a smooth curve.  Indeed,
one expects that as $M \rightarrow \infty$, $\rtherm \rightarrow M
\ravg + {\cal O}(\sqrt{M})$.  That is, a system of many hairpins should
exhibit self-averaging.  Moreover, because the limit of many hairpins
is essentially a thermodynamic limit, equivalence of ensembles must
also be
recovered. In fact, the
force-extension curve in the constant extension ensemble must
approach the disorder-averaged curve for the constant {\em force}
ensemble as $M$ becomes large.  In physical terms, there must be a constant tension along
the entire chain of hairpins; in the limit of many hairpins, each one
sees this tension rather than the extension imposed on the entire
chain.  Once there are enough competing hairpins, any equilibrium
experiment will give the same smooth curve.  Such smoothing, with its
attendant loss of structural information, has recently been observed
in simulations~\cite{terry-uli}.  Both the continuous increase of the
disorder average and the plateaus and jumps of a single hairpin can thus
appear in single molecule experiments.

We note in conclusion that the ideas from the physics of
one-dimensional disordered systems applied here to mechanical
denaturation experiments may find applications elsewhere in
biophysics.  To cite one example, the DNA-binding protein recA adheres
with a binding affinity that depends strongly on the nucleotide
sequencec~\cite{roy-albert}.  When ATP is replaced by the
non-hydrolyzable analog ATP-$\gamma\text{S}$, allowing the system to
reach equilibrium, the position of the point-like polymerization
boundary separating domains of polymerized recA from bare DNA can be
described by a coarse-grained model like Eqs.~\parenref{cale-defn}--\parenref{Z-defn}.  Similarly, the motion of a single boundary during
polymerization can be described as biased diffusion in a random force
field, and one might expect in appropriate parameter ranges to find strong disorder-induced slowing of the sort discussed in
section~\ref{dynamics}.  More
generally, the kinetics of multiple polymerization boundaries (associated
with multiple recA domains) on a single
long polynucleotide can naturally be mapped to the dynamics of kinks
in a one-dimensional random field Ising model, which is known to be
in the Sinai universality class~\cite{sinai,bouch-geo}.  Although the
relevance of such anomalous dynamics to the functioning of biological
systems {\em in vivo} remains to be established, these effects may
play a role in a number of {\em in vitro} assays.

It is a pleasure to thank Jean-Philippe Bouchaud, Ralf
Bundschuh, Daniel Branton, Daniel Fisher, Ulrich Gerland, and Terry Hwa for helpful
conversations and Roy Bar-Ziv and Albert Libchaber for introducing us
to recA polymerization.  We are also grateful to Rockefeller
University and to the Institute for
Theoretical Physics at the University of California, Santa Barbara,
for hospitality during the completion of this work.  This research
was supported by the NSF through grant DMR97-14725 and through the
Harvard MRSEC via grant DMR98-09363.  Work at UCSB was supported by
NSF grant PHY99-07949.

\appendix

\section{Simulation Methodology}
\label{sim-appendix}

This appendix describes the numerical method used to generate the data
points in Figures~\ref{four-polys}
through~\ref{rerand-fig}.  The simulations were performed on a
simplified model of dsDNA in which all A-T base pairs have a pairing energy
\epat\ and all G-C base pairs a pairing energy \epgc~\cite{heslot}.  In
contrast to the convention of subsection~\ref{semi-micro}, here we define pairing energies as the free energy difference between
the bound base pair and the two monomers subject to the tension $F$.
The average pairing energy of the sequence is thus $f$.  All base
pairs other than the $m$ unzipped bases are assumed to be closed, an
excellent approximation for dsDNA in physiological conditions.  We are
interested primarily in behavior near the unzipping transition, where
many bases have been unzipped.  In this regime, most of our
predictions depend only on universal properties of random walks, so
the simplifications in our model are justified.  Our results are always reported in terms of the
parameters $f$ and $\Delta$ that can be defined with reference only to
the large $m$ behavior of $\cale(m)$.  We assume for simplicity that A-T and G-C pairs
occur with equal probability $1/2$, and we take the pairing energies
to be $\epat = f - \sqrt{\Delta}$ and $\epgc = f + \sqrt{\Delta}$.
The disorder strength $\Delta$ is usually chosen to be between 1 and 9, while $f$
varies from $1$ down to a lower bound  determined by demanding
that $\mavg \approx \Delta/(2 f^2) \leq N/8$.  Here $N$ is the
total number of base pairs in the dsDNA, which we usually choose to
fall between $5 \times 10^5$ and $5 \times 10^6$.  For a given sequence $\{\varepsilon_i\}$, with each
$\varepsilon_i$ equal to either \epat\ or \epgc, $\cale(m)$ takes the
form $\cale(m) = \sum_{i=1}^m \varepsilon_i$.  The average and
variance of $\cale$ are then $\overline{\cale(m)} = m f$ and
$\overline{\cale(m)^2} - \overline{\cale(m)}^2 = \Delta m$, allowing
direct contact with the continuum limit described by
Eqs.~\parenref{cale-defn} and~\parenref{eta-defn}.  The temperature \kt\
is set to one. 

Our one-dimensional system is sufficiently simple that it is possible
to proceed by direct evaluation of the partition function $Z =
\sum_{m=0}^{N} \exp[-\cale(m)]$ and the average number of unzipped
bases $\mtherm = \sum_{m=0}^{N} m \exp[-\cale(m)]/Z$.  For each random sequence, successive values
of $\varepsilon_i$ are chosen at random, starting with
$\varepsilon_N$.  The running sums $Z_i \equiv \sum_{m=i}^N
\exp[-\cale(m) + \cale(i-1)]$ and $\mtherm_i \equiv \sum_{m=i}^N m
\exp[-\cale(m) + \cale(i-1)]$ are then
updated according to $Z_i = \exp(-\varepsilon_i) (1 + Z_{i+1})$ and $\mtherm_i =
\exp(-\varepsilon_i) (i + \mtherm_{i+1})$; once the sum is complete,
\mtherm\ is normalized by dividing by $Z$.  We keep separate sums for
each value of $f$, and, at each $i$, update each of them with the same
random choice of \epat\ or \epgc.  In some runs, we also kept track of
the running sum of $\varepsilon_i$ and of the location of the deepest
minimum encountered up to position $i$.  

The binned data in
Figures~\ref{mindistr-fig} and~\ref{jump-distr-fig}
represent the output of several thousand runs with independently
chosen random sequences and varying values of $\Delta$ and $N$.  In
Figure~\ref{mindistr-fig}, which plots the distribution of \mtherm, data points for
each value of $f$ from each run were rescaled appropriately and used
together to construct the histogram.  Similarly, all pairs of points
with $f_2/f_1 \approx 0.77$ were rescaled and used in making the
histogram of \dm\ in Figure~\ref{jump-distr-fig}; in order to account for the
predicted delta function at $\dm = 0$, a fraction $f_2/f_1$ of the
total number of data points was subtracted from the number of counts
in the bin that included $\dm = 0$.

\bfig

\epsffile{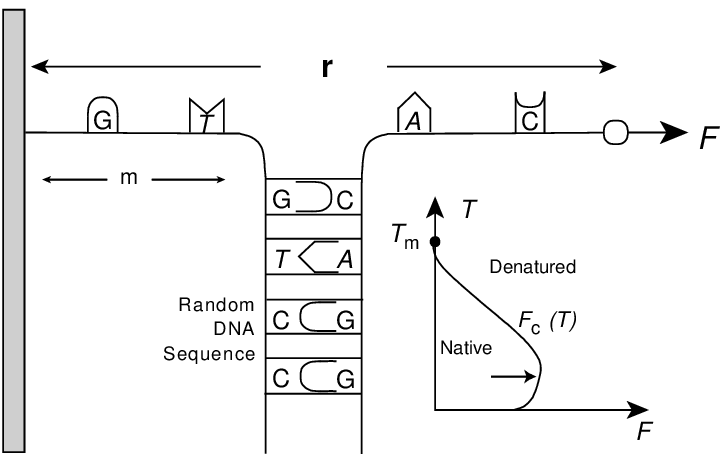}

\caption{Sketch of the DNA unzipping experiment.  One of the single
strands of a dsDNA molecule with a random base sequence is attached by
its end to a solid surface, and the other is pulled away from the
surface with a constant force \fb.  As a result, the double strand
partially denatures, separating $m$ base pairs ($m = 2$ in the
figure).  The distance between the ends of the two single strands, or
{\em extension}, is $\rb$. {\em Inset}: Schematic phase diagram in the
temperature-pulling force ($T$-$F$) plane of a dsDNA molecule in three
dimensions.  At low enough $T$ and $F$, the polymer is in the native,
double-stranded phase.  At the phase transition line $\fc(T)$, the DNA
denatures and the two strands separate.  Thermally-induced {\em
melting} occurs at zero force at a temperature $T_{\text{m}}$.  As
indicated by the arrow, this paper considers instead the {\em
unzipping} transition, in which the phase transition line is crossed
at non-zero $F$.  The reentrance at low temperatures is predicted
in~\protect\cite{maritan1}.
\label{fig1}}
\efig

\newpage

\bfig

\epsffile{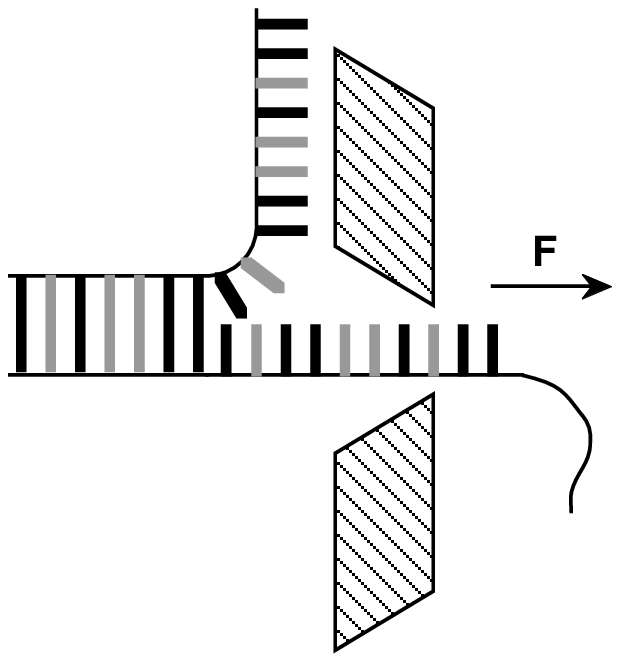}

\caption{Schematic of dsDNA unzipping through a narrow
pore~\protect\cite{dan-personal}.  The pore is assumed to be large
enough that single-stranded DNA, but not double-stranded DNA, can fit
through it.  Under the influence of an electric field or comparable
force \fb, one single strand inserts into the channel and is
gradually pulled through.  As the strand is drawn through the
pore, it must unzip from its complementary strand.
\label{pore-unzip}}
\efig

\bfig

\epsffile{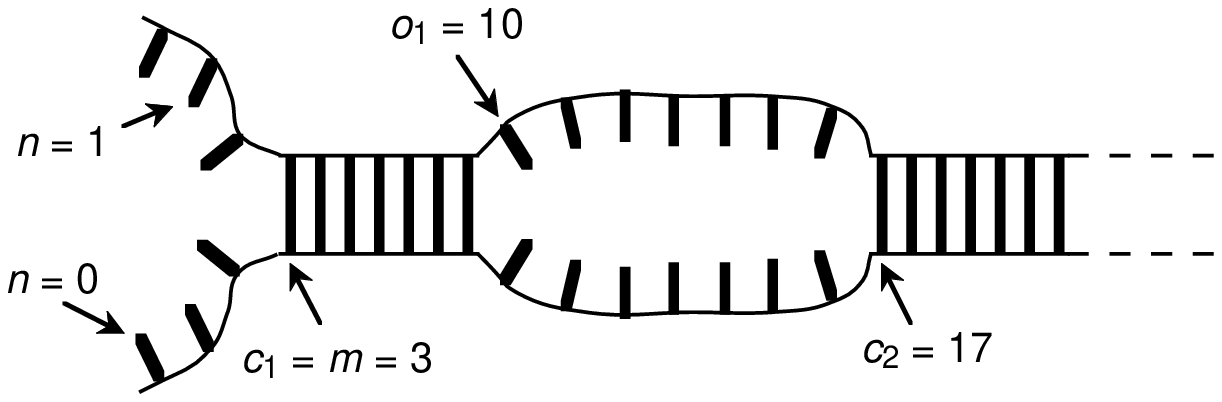}

\caption{Definition of the variables $c_i$ and $o_i$ in the Ising-like
model [Eq.~(\protect\ref{hi})].  In this figure, 3 bases are open at the end of the dsDNA.
Counting the first open base as $n=0$, the location of the first
closed base is then $c_1 = 3$.  Similarly, the next open base is at
$o_1 = 10$.
\label{ising-fig}}

\efig

\bfig

\epsffile{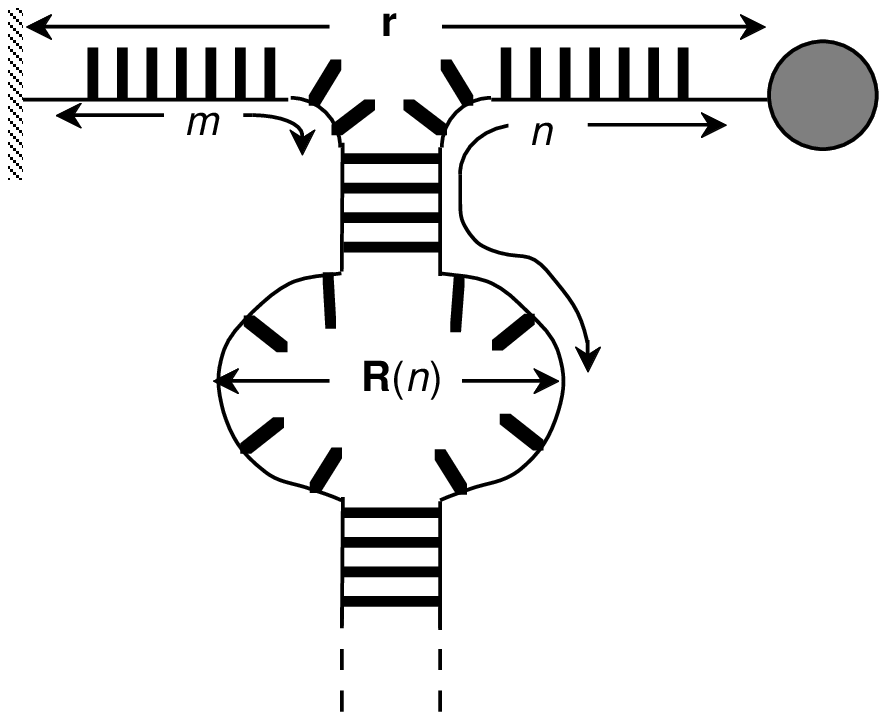}

\caption{Definition of the variables in the continuum model [Eq.~(\protect\ref{hc})].  The
distance between the ends of the two single strands (the {\em
extension}) is $\rb$, and the number of open bases is $m$.  The bases
are indexed by $n$; the separation between the two single strands at the
$n^{\text{th}}$ base pair from the end is given by $\rlb(n)$.
\label{var-defn}}

\efig

\bfig

\epsffile{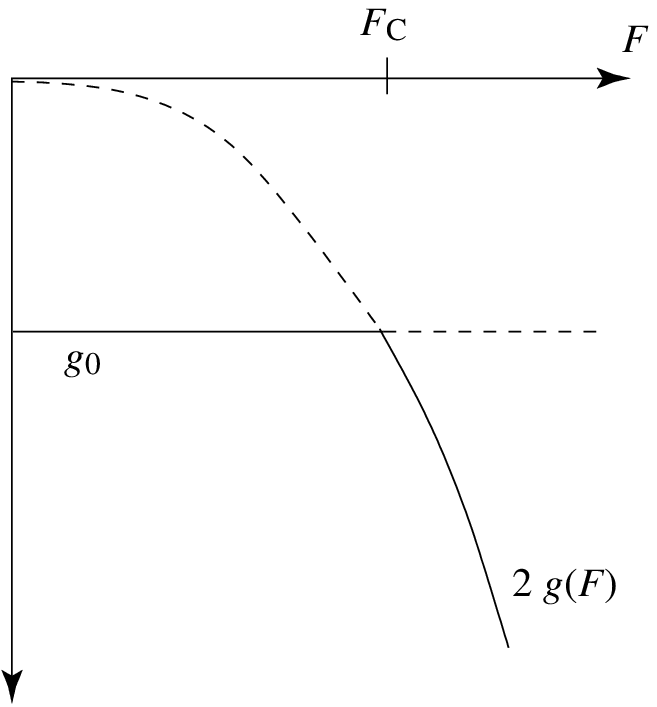}

\caption{Sketch of the bulk free energies per base pair $g_0$ of the zipped phase and
$2 g(F)$ of the unzipped phase as a function of the applied force
$F$.  These negative energies are measured relative to the free energy
of a base pair at infinite separation with $F = 0$.  While $g_0$ is independent of $F$, $2 g(F)$ decreases with
increasing $F$.  At a critical force value \fc, the zipped phase
becomes unstable relative to the unzipped phase, and a
phase transition occurs.  The equilibrium free energy per base pair as
a function of $F$ is given by the solid curves; the discontinuous
change in slope at \fc\ indicates a first order transition.
\label{first-order} }

\efig

\bfig

\epsffile{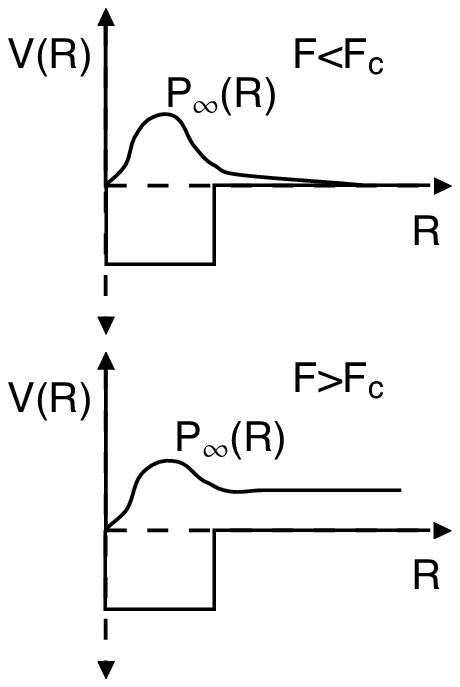}

\caption{Schematic of the base separation probability $P_{\infty}(R)$
below and above the first order unzipping transition (the probability
is expected to depend only on the radial distance $R$, not on angular coordinates).  Below
the transition, $P_{\infty}(R)$ decays quickly to zero beyond the
range of the attractive potential $V(\rlb)$.  Above the transition, in
contrast, it approaches a constant non-zero value as $R \rightarrow \infty$.
\label{radial-distr} }

\efig

\bfig

\epsffile{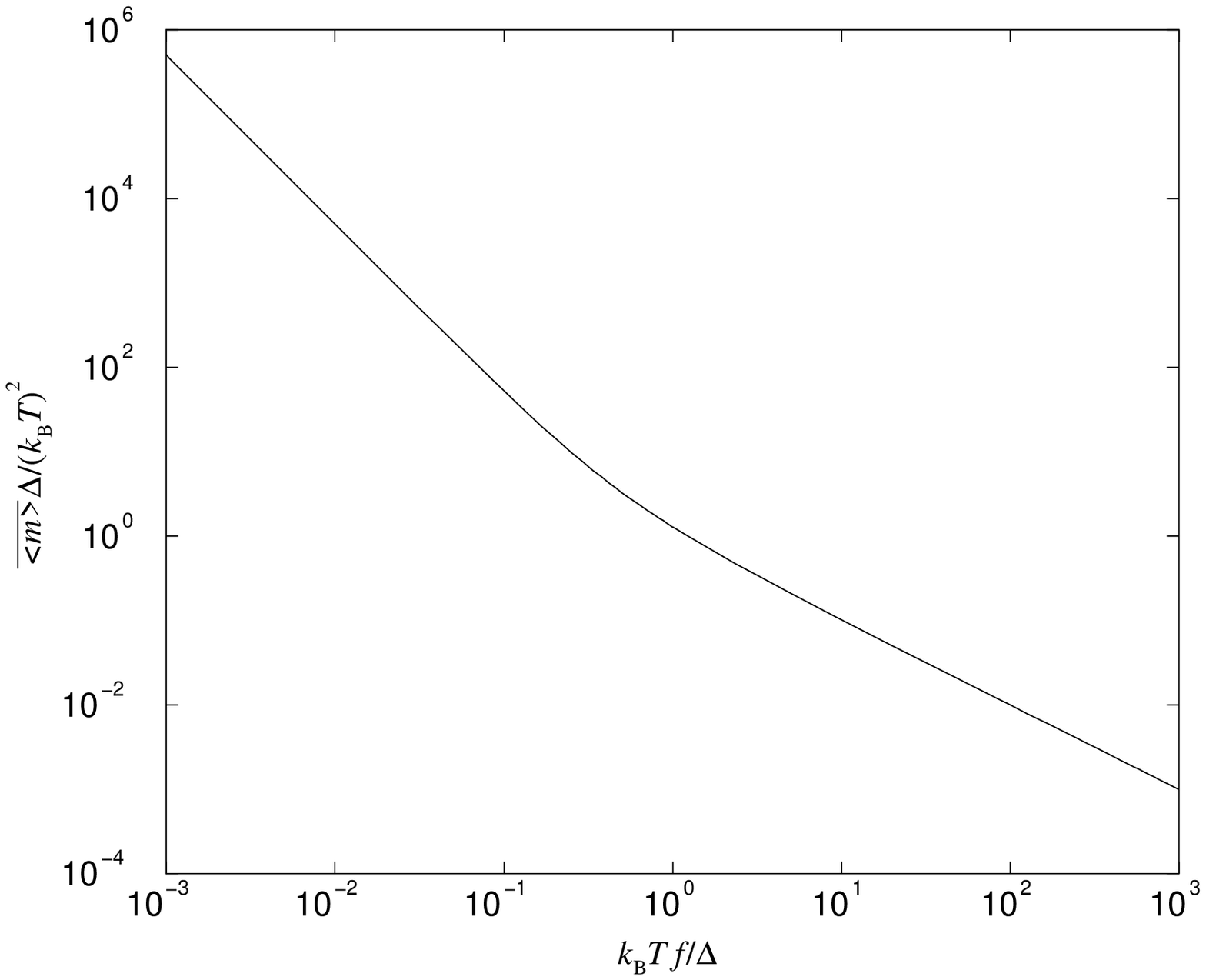}

\caption{Log-log plot of Eq.~(\protect\ref{mavg})
for \mavg\ as a function of $f = 2 g(F) - g_0 \sim \fc - F$.  For large $f$, the plot has slope
$-1$, but it crosses over to slope $-2$ at $f \approx \Delta/\kt$.
\label{crossover}}
\efig

\bfig

\epsffile{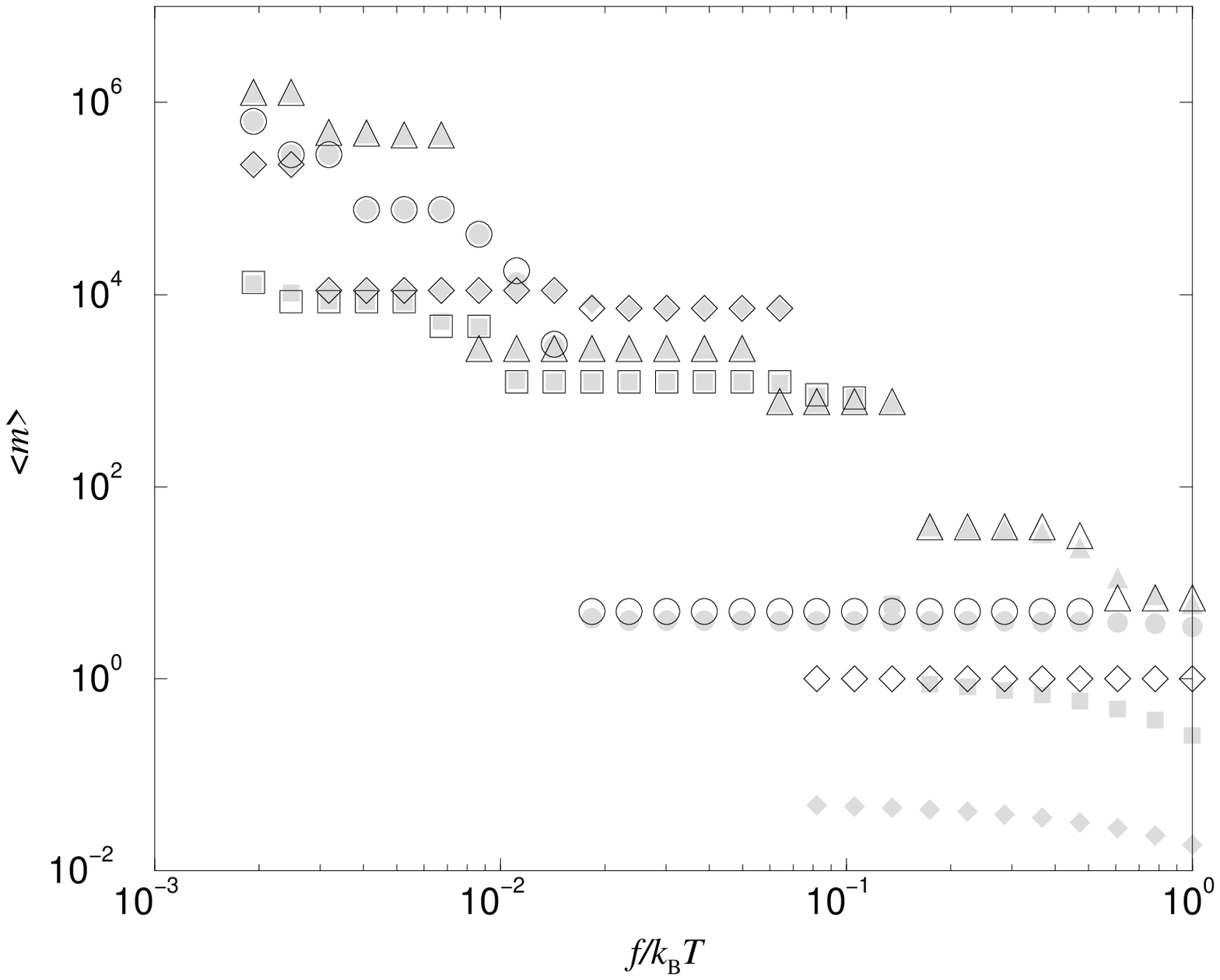}

\caption{Log-log plot of the average number of bases opened \mtherm\
(closed symbols) and the location of the absolute minimum \mmin\ of
$\cale(m)$ (open symbols) as a function of the distance $f$ from the
unzipping transition.  Both variables are plotted for each of four
individual polymers, represented by four different symbol shapes, with
independently chosen random sequences (variance $\Delta = 9 (\kt)^2$) of length $N = 5\times 10^6$ bases.
Note that, except when $m = {\cal O}(1)$, \mtherm\ and \mmin\ coincide
very well.  The energy landscapes for the four duplexes are plotted
for a particular value of $f$ in Fig.~\protect\ref{four-ens}.
\label{four-polys}}
\efig

\bfig

\epsffile{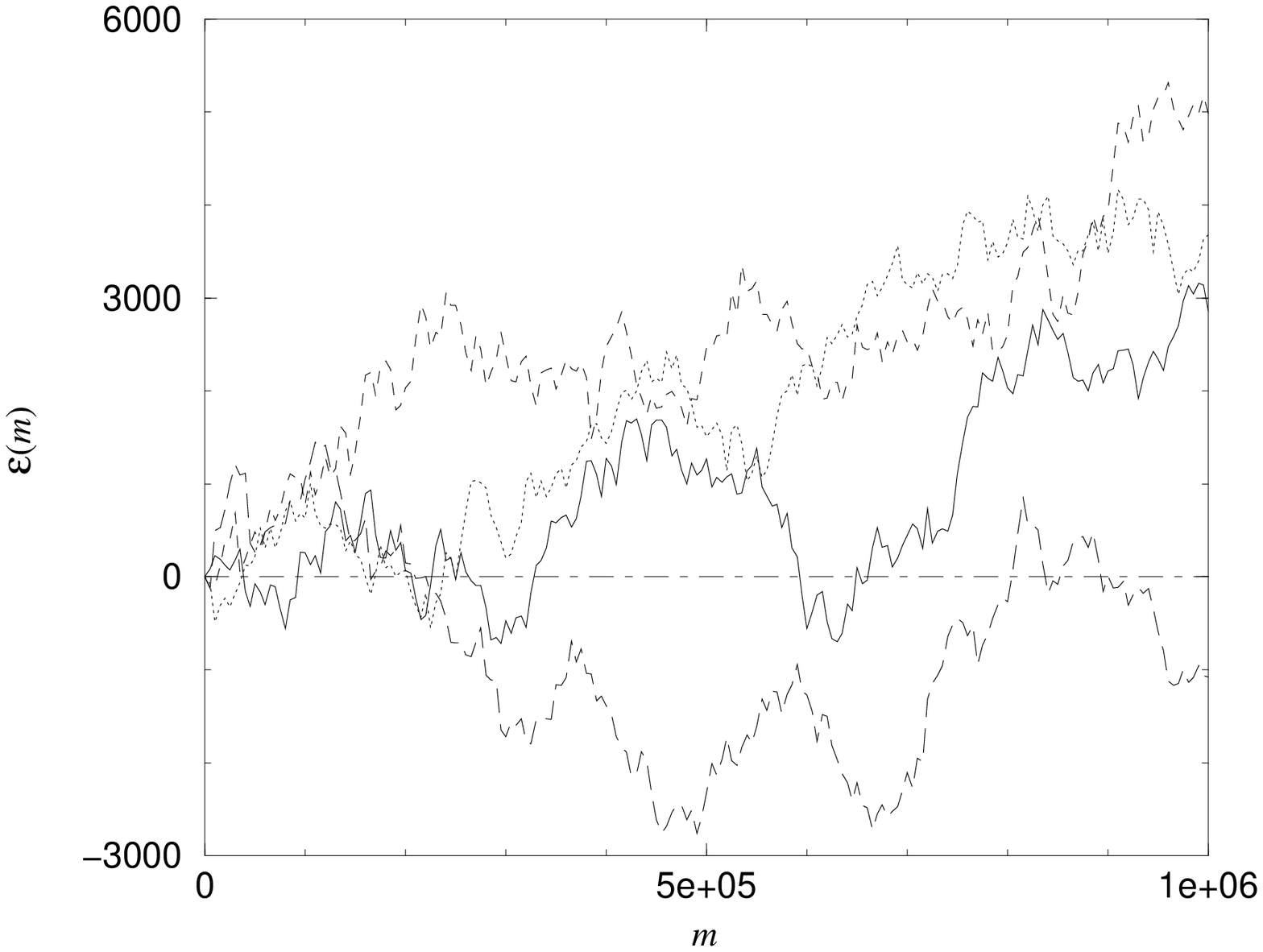}

\caption{Plot of four different random realizations of $\cale(m)$.
All four random walks have the same variance $\Delta = 9 (\kt)^2$ and average bias
$f = 0.0025 \kt$.  All four also pass below $\cale = 0$,
suggesting that, near the unzipping transition, dsDNA molecules with
random sequences will usually have {\em energetic} reasons to
partially unzip.  The four energy landscapes are taken from the four
polymers whose force-extension curves are shown in
Figure~\protect\ref{four-polys}; the solid, dashed, dotted, and
long-dashed curves correspond, respectively, to the circles, squares,
diamonds, and triangles.
In order to focus on regions where $\cale(m)$ is near zero, the
landscapes for $m > 10^6$ are not shown.
\label{four-ens}}
\efig

\bfig

\epsffile{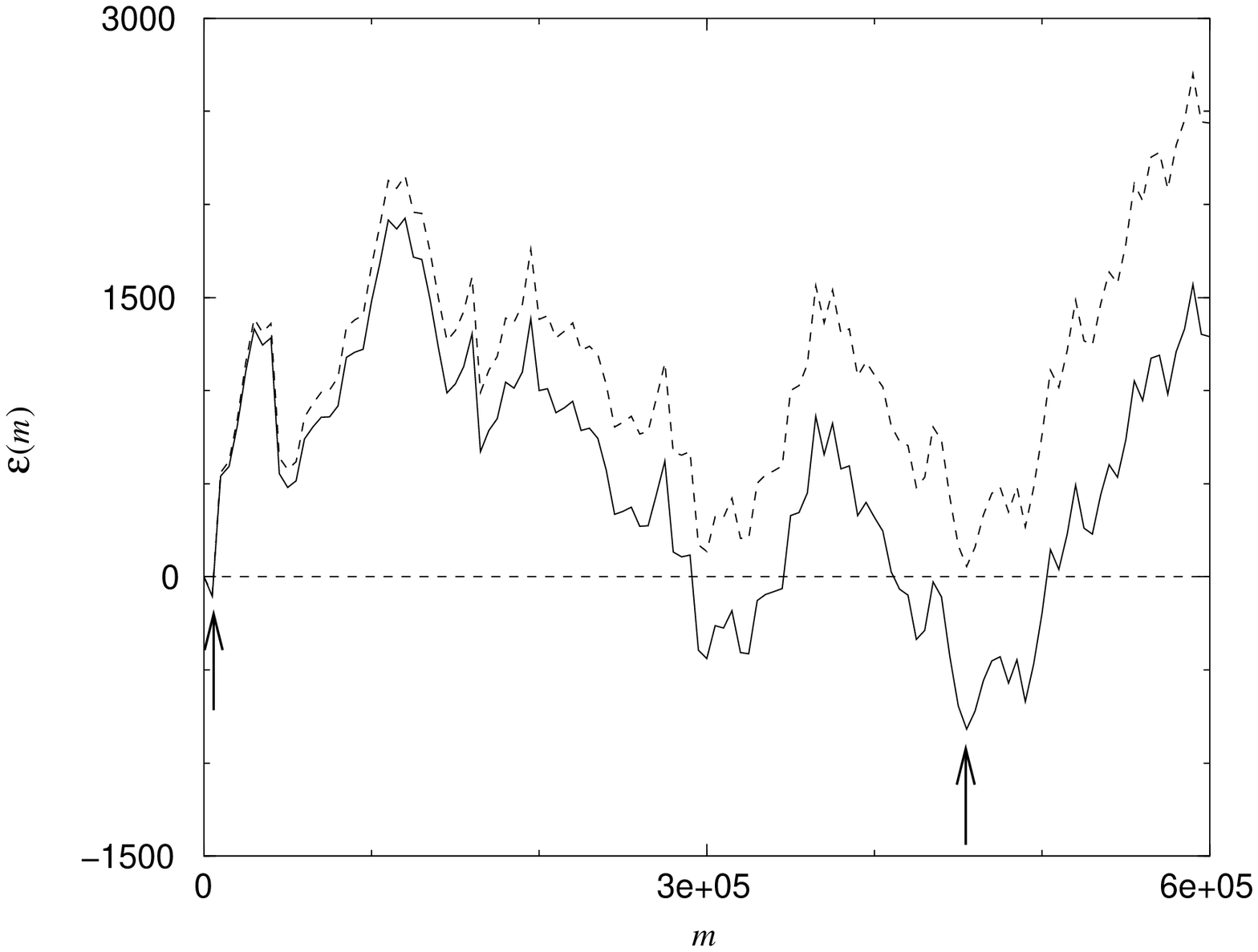}

\caption{Plot illustrating the physical origin of jumps in extension
versus force during unzipping.  The two curves represent random walks
$\cale(m)$ with identical random contributions $W(m)$, but different
average biases $f_1 = 0.0087$ (upper curve) and $f_2 = 0.0067$ (lower
curve).  As indicated by the arrows, in
the upper curve, the absolute minimum \mmin\ is at $\mmin \approx
5,000$, while in the lower curve, it is at $\mmin \approx 445,000$.  As
$f$ is tuned from $f_1$ down to $f_2$, \mmin\, and thus \mtherm\, jump
from one minimum to the other.
\label{energy-tilt}}
\efig

\bfig

\epsffile{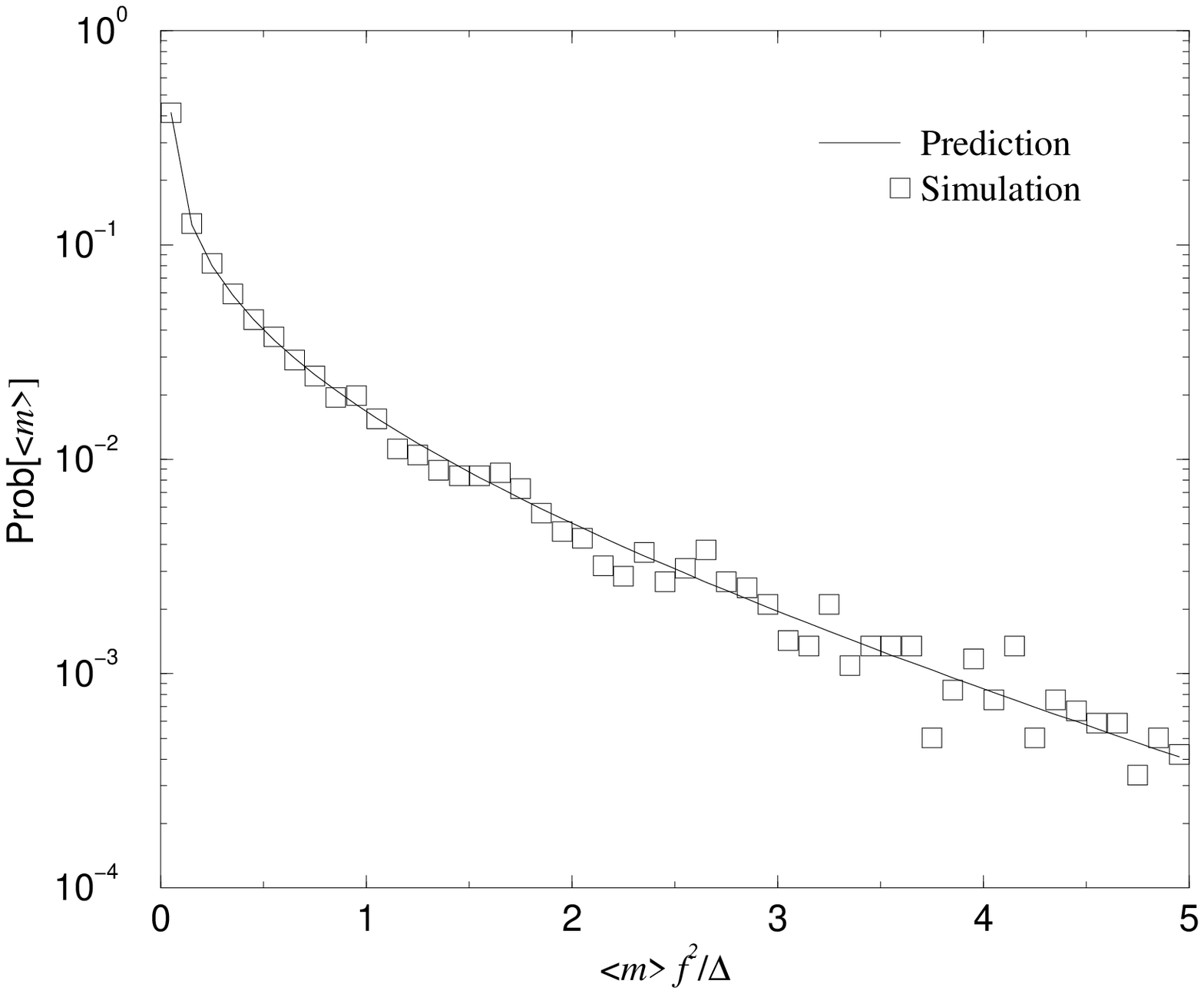}



\caption{Log-linear plot of the distribution over different random
sequences of the average number of
opened bases \mtherm.  The horizontal
axis gives \mtherm, suitably rescaled so that random sequences with
different values of $f$ and $\Delta$ can be compared.  The vertical
axis shows
the log of the probability of seeing a particular \mtherm.  The squares
represent binned data from numerical simulations (described in the
Appendix), the solid curve the analytic prediction of Eq.~(\protect\ref{min-distr-eqn}) based on the
assumption that $\mtherm = \mmin$.  This prediction has no adjustable
parameters.  The scatter seen for large $\mtherm
f^2/\Delta$ is the result of counting noise.
\label{mindistr-fig}}
\efig

\bfig

\epsffile{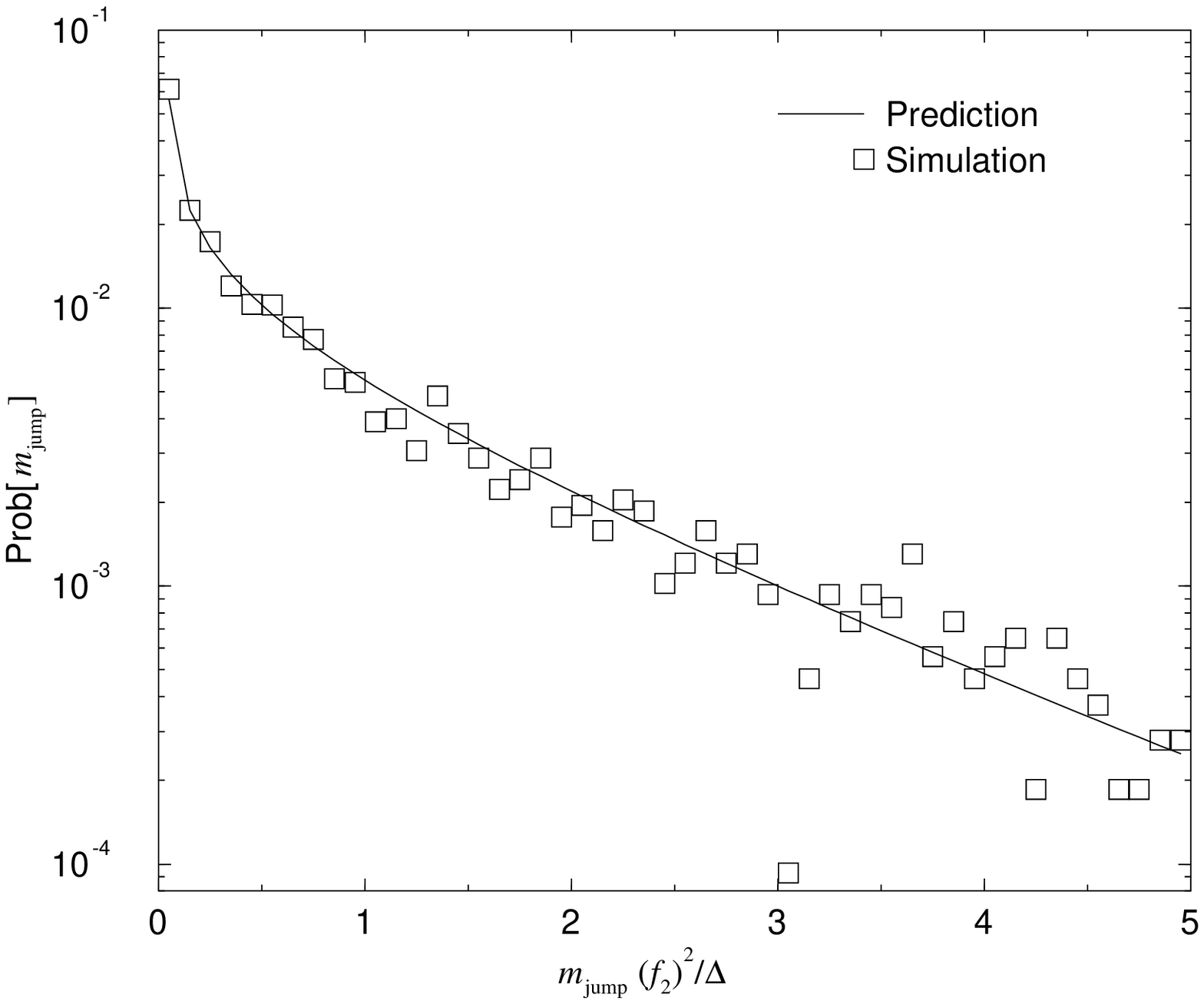}

\caption{Log-linear plot of the distribution of jumps \dm\ for
$f_2/f_1 \approx 0.77$.  $\dm f_2^2/\Delta$ is plotted on the
horizontal axis, the log of the probability of \dm\ on the vertical
axis.  The points
represent binned data from numerical simulations (described in the
Appendix), the solid curve an analytic prediction (no adjustable parameters) based on the
assumption that $\mtherm = \mmin$.  The scatter seen for large $\dm
f_2^2/\Delta$ is the result of counting noise.
\label{jump-distr-fig}}
\efig

\bfig

\epsffile{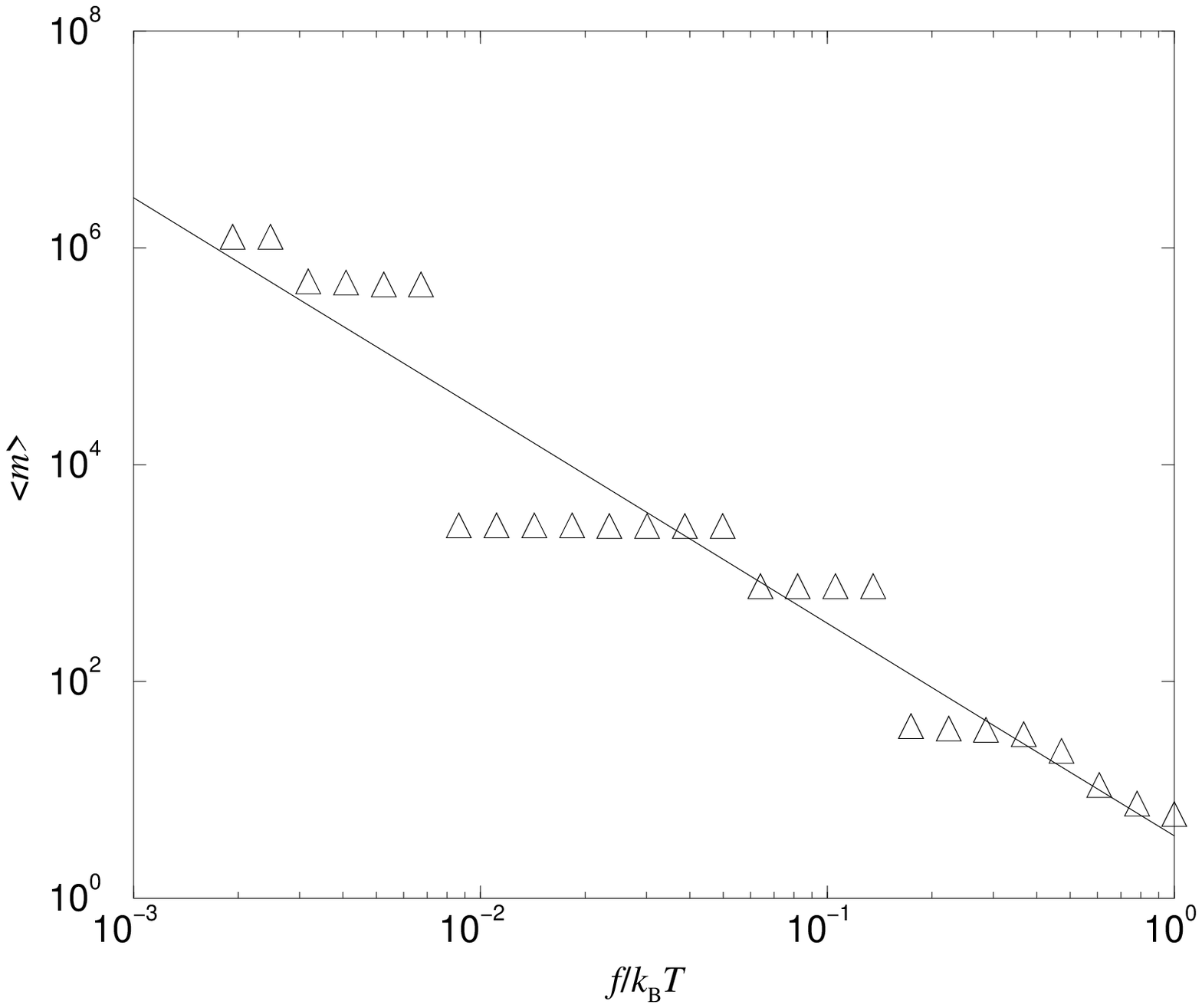}

\caption{Plot illustrating the recovery of the disorder-averaged
scaling law $\mavg \simeq \Delta/2 f^2$ in the force-extension curve
of a single random heteropolymer.  The points give \mtherm\ as a
function of $f$ for a single polymer; the solid line is the best-fit
power law, with exponent $-1.96 \pm 0.12$.
\label{rerand-fig}}
\efig

\bfig

\epsffile{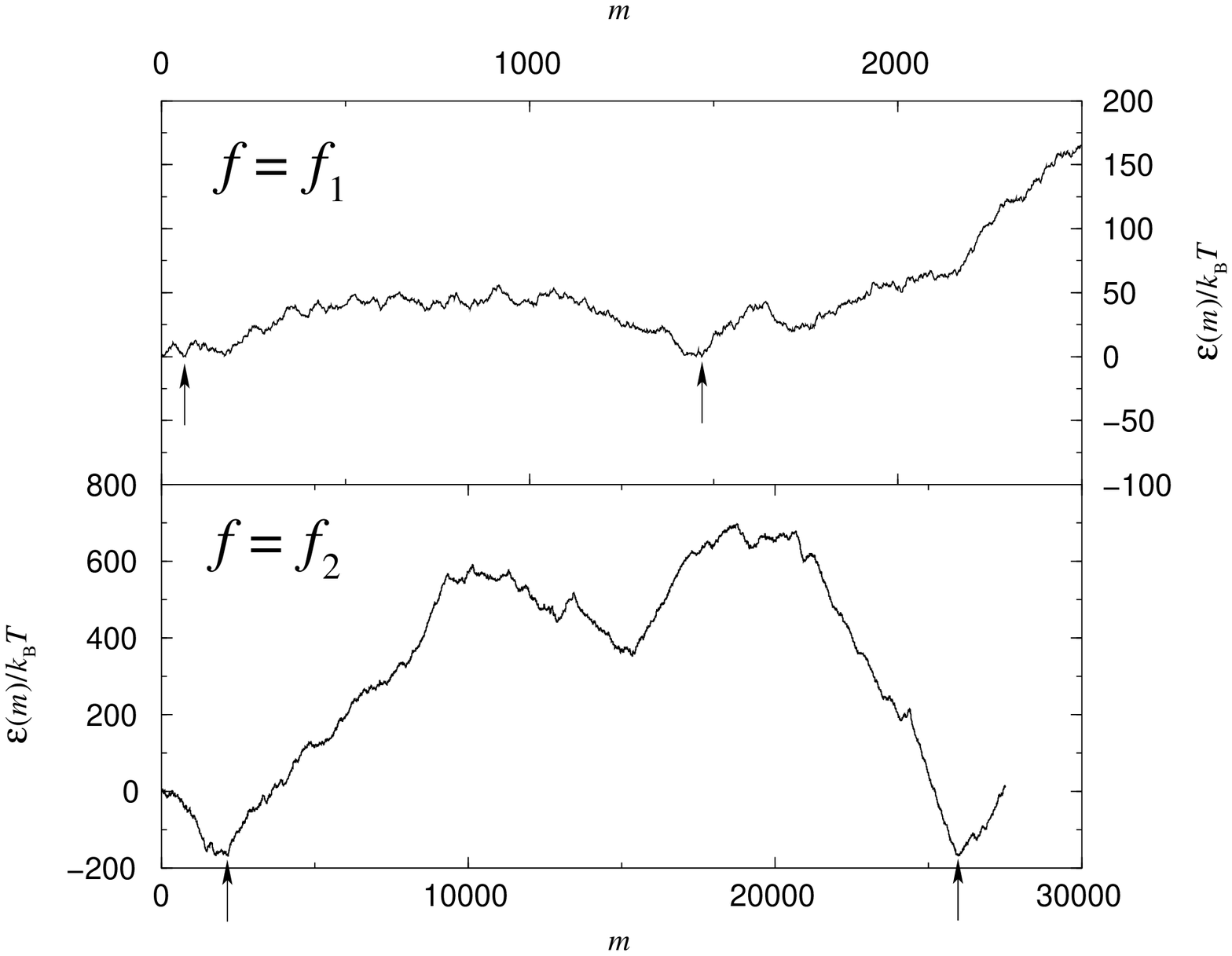}

\caption{The energy landscape $\cale(m)$ for unzipping bacteriophage lambda
DNA at two different biases.  In this figure, the base pairs are opened in the
reverse of the conventional~\protect\cite{gen-bank} order, starting with base
number 48502.  Base pairing and stacking energies are taken
from~\protect\cite{dna-en} and are scaled by $\kt$, with $T =
37^{\circ}\text{C} = 310 \text{K}$.  The biases $f_1$ and $f_2$ are the
locations of the two jumps marked in the force-extension curve of
Figure~\protect\ref{lam-opening}.  The locations of the two minima
that exchange stability at each bias are indicated by arrows.  Note
the difference in scales between the upper and lower plots.
\label{lam-en}}
\efig

\bfig

\epsffile{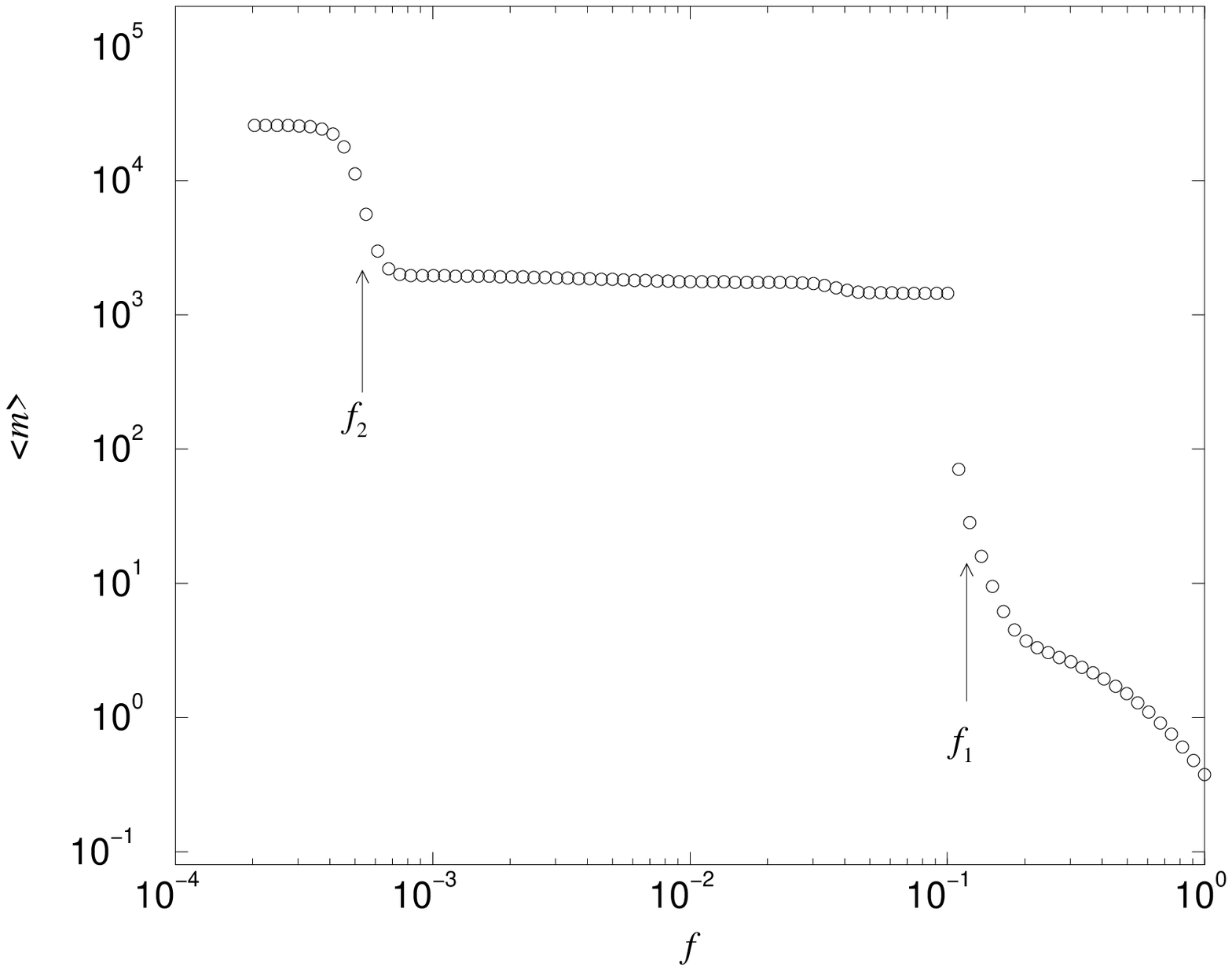}

\caption{Log-log plot of the average number of open bases \mtherm\ versus
bias $f$ for unzipping bacteriophage lambda DNA.  The energy
$\cale(m)$ is as in Fig.~\protect\ref{lam-en}.  The three plateaus
correspond to the minima of $\cale(m)$ at $m \approx 1$, $m \approx
1500$, and $m \approx 26,000$; the jumps between them occur at biases
$f_1$ and $f_2$ as indicated in the figure. Assuming freely jointed
chain elasticity for ssDNA [Eq.~(\protect\ref{fjc})], with $b = 1.5
\text{nm}$~\protect\cite{busta}, the definition of $f$
[Eq.~(\protect\ref{f-defn})] implies that these biases correspond respectively
to forces of $F_1 = 7.90 \text{pN}$ and $F_2 = 8.14 \text{pN}$.  The
middle plateau is actually subdivided into three smaller plateaus,
separated by jumps between nearby minima.  Similarly, a local minimum
at $m = 60$ is the most stable for a small range of $f$ between the
plateaus at $m=1$ and $m \approx 1500$.
\label{lam-opening}}
\efig

\bfig

\epsffile{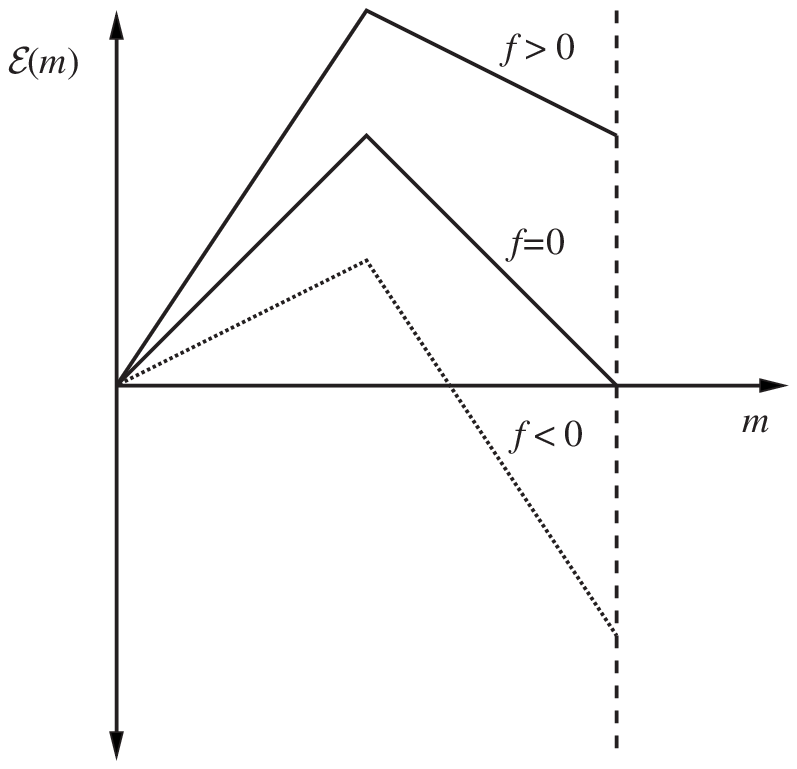}

\caption{Schematic energy landscape for a designed oligonucleotide
duplex that could be used to measure base pairing and stacking
energies.  The duplex is chosen to have stronger base pairs near the
end from which is it opened, and weaker base pairs at the far end.  The
energy of opening $\cale(m)$ thus first slopes upwards, then
downwards, and the only two minima occur for a completely unzipped and
completely zipped ($m=0$) duplex.  As the bias is tuned through the
unzipping transition, the two minima exchange stability, giving rise
to a sharp unzipping transiton (see Fig.~\protect\ref{enmeas-open}).
\label{enmeas-tilt} }

\efig

\bfig

\epsffile{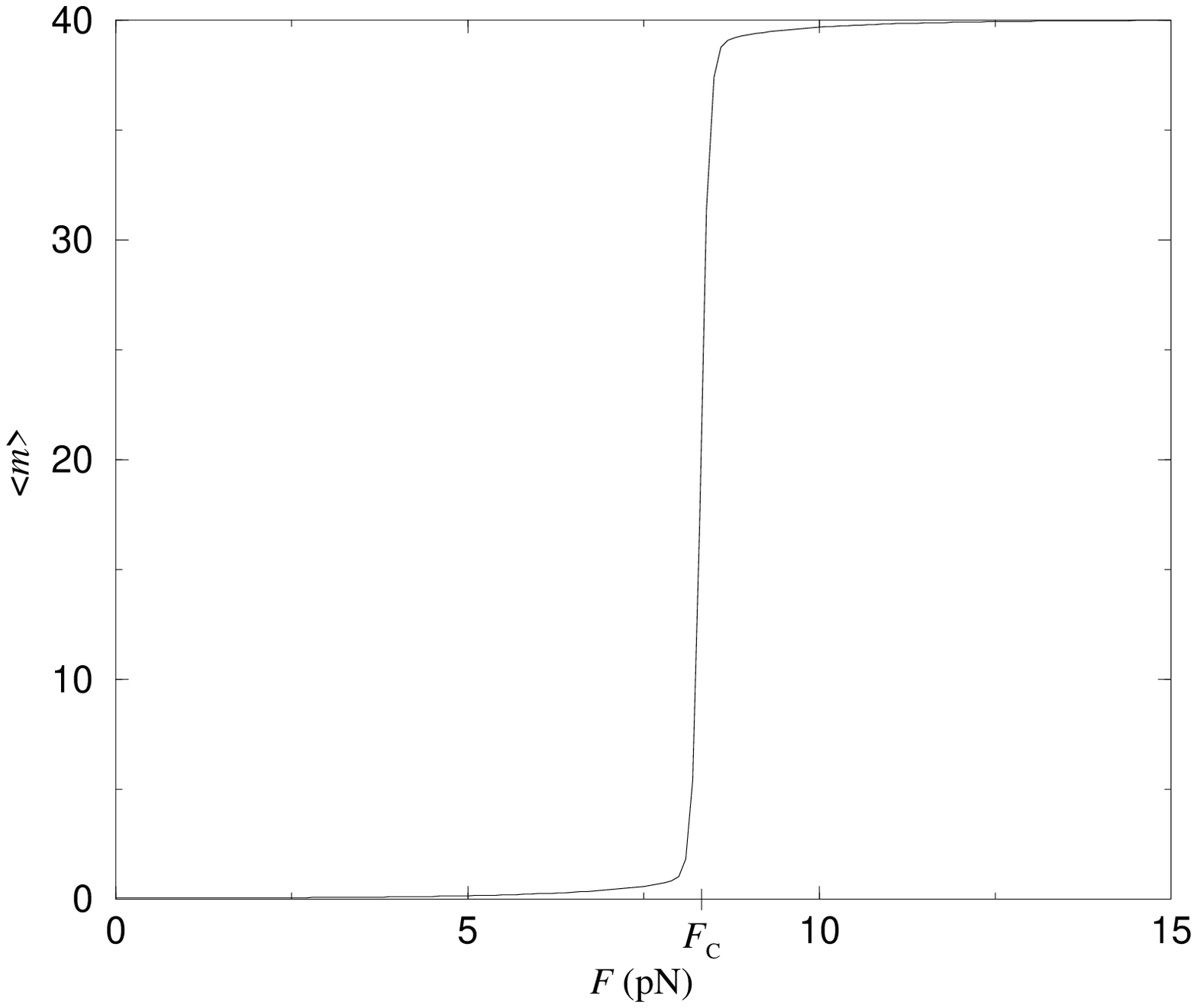}

\caption{Force extension plot for a designed oligonucleotide duplex
that could be used to measure base pairing and stacking energies (see
Fig.~\protect\ref{enmeas-tilt}).  Starting from the end from which it
is being unzipped, the duplex has sequence
$5'(\text{A})_{20}(\text{C})_{20}3'$, with base pairing energies taken
from~\protect\cite{dna-en}. The sharp
unzipping transition allows an accurate measurement of $\fc = 8.32
\text{pN}$, and thus of the energies stabilizing the duplex.  Forces
are calculated assuming that ssDNA is a freely jointed chain
[Eq.~(\protect\ref{fjc})], with Kuhn length $b = 1.5 \text{nm}$~\protect\cite{busta}.
\label{enmeas-open}}

\efig

\newpage

\bfig

\epsffile{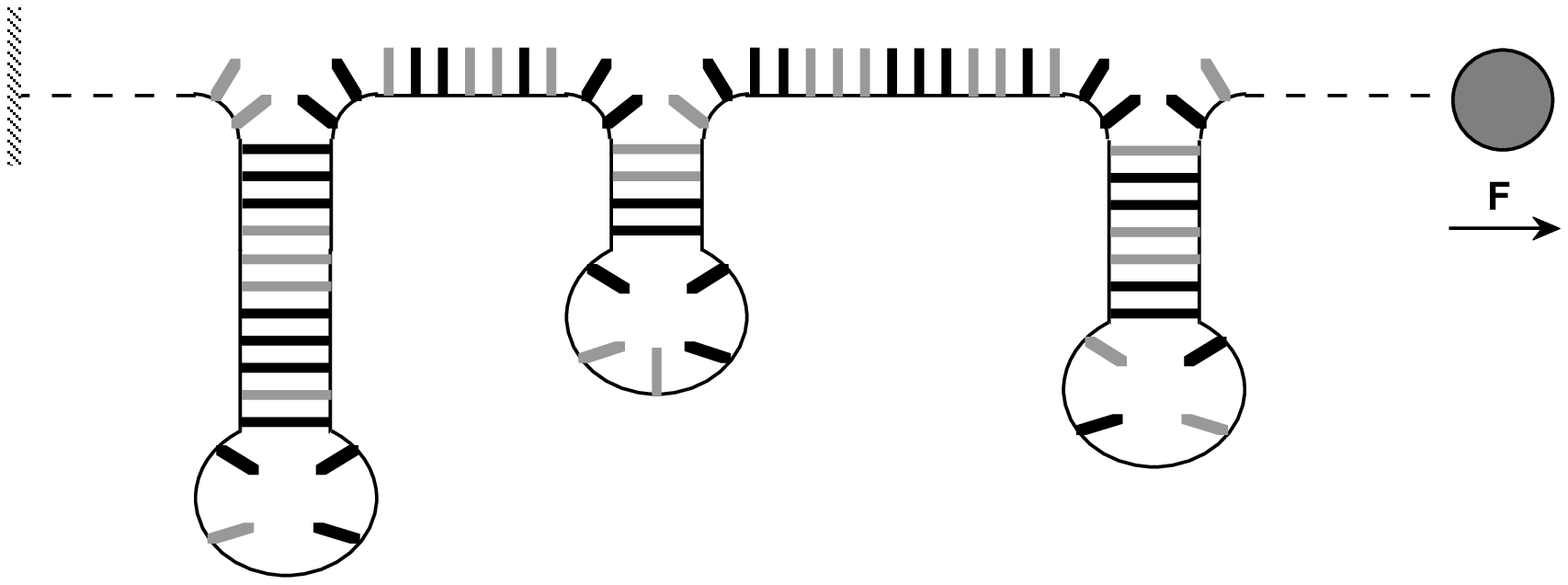}

\caption{Sketch of several RNA stems being opened in parallel, as
might occur in a micromechanical experiment on a ribozyme or other
folded RNA molecule.  If each stem has an independently chosen random
sequence, then in the limit of a large number of long stems, the number of
unzipped bases will equal the disorder averaged value \mavg.  The
measured force-extension curve must then be smooth and monotonic in
any ensemble.
\label{rna-fig}}
\efig


\begin{thebibliography}{99}
\bibitem{sim-note} Simulations consisted of exact evaluation of the
partition function for randomly generated sequences; details are given
in appendix~\ref{sim-appendix}.
\end{thebibliography}
\end{document}